\documentclass[fleqn]{2020SCGE}
\usepackage{color}
\usepackage{float}
\usepackage{natbib}
\usepackage{gensymb}
\usepackage{graphicx}
\usepackage{tablefootnote}
\usepackage[toc]{multitoc}
\usepackage{rotating}
\usepackage{lscape}
\usepackage{pdflscape}
\usepackage{setspace}
\usepackage{url}
\usepackage{hyperref}

\newcommand{\blue}[1]{{\leavevmode\color{blue} #1}}
\newcommand{\kms}{${\rm km~s}^{-1}$}
\newcommand{\HII}{H\,{\footnotesize{II}} }
\newcommand{\HI}{H\,{\footnotesize{I}} }

\newcommand{\mjyb}{{\rm mJy~beam}^{-1}}

\newcommand{\mjybkms}{{\rm mJy~km~s}^{-1}}
\newcommand{\jybkms}{{\rm Jy~km~s}^{-1}}

\newcommand{\dotsec}{\rlap.{''}}
\newcommand{\dotmin}{\rlap.{'}}
\newcommand{\dotdeg}{\rlap.{^\circ}}

\newcommand{\msun}{M_{\odot}}

\newcommand{\arcmin}{$^{\prime}$}

\begin{document}

\ensubject{subject}

%%%%%%%%%%%%%%%%%%%%%%%%%%%%%%%%%%%%%%%%%%%%%%%%%%%%%%%
%%% Authors do not modify the information below
%%% ????????????????
%%% ??????????, ????????????{}, ???????????????????
%Letter to the Editor??Article%??????
\ArticleType{Article}%??Article
%\SpecialTopic{SPECIAL TOPIC: }%???????
\Year{2024}
\Month{January}
\Vol{67}
\No{1}
\DOI{10.1007/s11433-023-2219-7}
\ArtNo{219511}
\ReceiveDate{June 20, 2023}
\AcceptDate{September 8, 2023}
\OnlineDate{December 8, 2023}
%%%%%%%%%%%%%%%%%%%%%%%%%%%%%%%%%%%%%%%%%%%%%%%%%%%%%%%

%%% title: ????
%%%   \title{title}{title for citation}
\title{The FAST All Sky H{\,\Large{I}} Survey (FASHI): the first release of catalog}{FASHI}

%%% Corresponding author: ???????
%%%   \author[number]{Full name}{{email@xxx.com}}
%%% General author: ???????
%%%   \author[number]{Full name}{}
\author[1,2,\footnote{Corresponding authors (C.-P. Zhang, email: \url{cpzhang@nao.cas.cn}; M. Zhu, email: \url{mz@nao.cas.cn}; P. Jiang, email: \url{pjiang@nao.cas.cn})}]{Chuan-Peng Zhang}{}%
\author[1,2,$^\blue\ast$]{Ming Zhu}{}%
\author[1,2,$^\blue\ast$]{Peng Jiang}{}%
\author[3]{Cheng Cheng}{}%
\author[4]{Jing Wang}{}%
\author[2]{Jie Wang}{}%
\author[1,2]{\\Jin-Long Xu}{}%
\author[1,2]{Xiao-Lan Liu}{}%
\author[1,2]{Nai-Ping Yu}{}%
\author[1,2]{Lei Qian}{}%
\author[2]{Haiyang Yu}{}%
\author[1,2]{Mei Ai}{}%
\author[2]{Yingjie Jing}{}%
\author[2]{\\Chen Xu}{}%
\author[2]{Ziming Liu}{}%
\author[1,2]{Xin Guan}{}%
\author[1,2]{Chun Sun}{}%
\author[1,2]{Qingliang Yang}{}%
\author[1,2]{Menglin Huang}{}%
\author[1,2]{\\Qiaoli Hao}{}%
\author[1,2]{FAST Collaboration}{}%
%%% Author information for page head. ?ü?е????????
%%% ??????????????, ??????????author???
\AuthorMark{Zhang C.-P.}%\authorcr????????

%%% Authors for citation. ????????е????????
%%% ??????????????, ??????????author???
\AuthorCitation{Zhang C.-P., et al}

%%% Address. ???
%%%   \address[number]{Address, City {\rm Postcode}, Country}
\address[1]{Guizhou Radio Astronomical Observatory, Guizhou University, Guiyang 550000, People's Republic of China}
\address[2]{National Astronomical Observatories, Chinese Academy of Sciences, Beijing 100101, People's Republic of China}
\address[3]{Chinese Academy of Sciences South America Center for Astronomy, National Astronomical Observatories, CAS, Beijing 100101, China} 
\address[4]{Kavli Institute for Astronomy and Astrophysics, Peking University, Beijing 100871, People's Republic of China}
%\contributions{}%????????

%%% Abstract. ??
\abstract{The \textbf{F}AST \textbf{A}ll \textbf{S}ky \textbf{H\,{\footnotesize{I}}} survey (FASHI) was designed to cover the entire sky observable by the Five-hundred-meter Aperture Spherical radio Telescope (FAST), spanning approximately 22000 square degrees of declination between $-14\degree$ and $+66\degree$, and in the frequency range of 1050-1450\,MHz, with the expectation of eventually detecting more than 100000 \HI sources. Between August 2020 and June 2023, FASHI had covered more than 7600 square degrees, which is approximately 35\% of the total sky observable by FAST. It has a median detection sensitivity of around 0.76\,$\mjyb$ and a spectral line velocity resolution of $\sim$6.4\,\kms~at a frequency of $\sim$1.4\,GHz. As of now, a total of 41741 extragalactic \HI sources have been detected in the frequency range 1305.5-1419.5\,MHz, corresponding to a redshift limit of $z\lesssim0.09$. By cross-matching FASHI sources with the Siena Galaxy Atlas (SGA) and the Sloan Digital Sky Survey (SDSS) catalogs, we found that 16972 (40.7\%) sources have spectroscopic redshifts and 10975 (26.3\%) sources have only photometric redshifts. Most of the remaining 13794 (33.0\%) \HI sources are located in the direction of the Galactic plane, making their optical counterparts difficult to identify due to high extinction or high contamination of Galactic stellar sources. Based on current survey results, the FASHI survey is an unprecedented blind extragalactic \HI survey. It has higher spectral and spatial resolution and broader coverage than the Arecibo Legacy Fast ALFA Survey (ALFALFA). When completed, FASHI will provide the largest extragalactic \HI catalog and an objective view of \HI content and large-scale structure in the local universe.  
% \Authorfootnote \noindent 
}

%%% Keywords. ?????

\maketitle

\begin{multicols}{2}
%%%%%%%%%%%%%%%%%%%%%%%%%%%%%%%%%%%%%%%%%%%%%%%%%%%%%%%%%%%%
%% Text of article.
%%%%%%%%%%%%%%%%%%%%%%%%%%%%%%%%%%%%%%%%%%%%%%%%%%%%%%%%%%%%
\section{Introduction}    %% first-level sections will be auto-capitalized
\label{sect:intro}

Neutral hydrogen (H\,{\footnotesize{I}}) is a significant component of the interstellar medium within disk galaxies. The measurement of its abundance and kinematics, via the 21\,cm emission line, has been employed to address numerous astrophysical issues, including studies on the large-scale distribution of galaxies and their secular evolution \citep[e.g.,][]{Wong2006,Cheng2020,Xu2022}. The line emission from \HI provides crucial information about the distribution of neutral gas and its velocity field within a galaxy \citep[e.g.,][]{Springob2005}. \HI line offers a way to obtain redshifts, gas masses, and rotational widths for standard galaxies, tracing the evolution of tidal phenomena, and furnish quantitative assessment of the potential for future star formation through \HI content. Moreover, because of its relatively simple physics, the \HI line provides a valuable means of tracing cool gas mass and star formation potential in nearby galaxies, and enables the investigation of low-luminosity, gas-rich object populations. Low surface brightness, dwarf, and gas-rich galaxy populations are often underrepresented in optical or infrared surveys. Although public wide-area optical or infrared surveys and their associated spectroscopic surveys are good at detecting luminous ellipticals, bright spirals, and bursting or active galaxies, they are substantially less complete in tracing the low surface brightness, dwarf, and gas-rich galaxy populations which actually dominate the local population. Thus, large-scale extragalactic \HI surveys remain necessary to take stock of gas-rich galaxies in the proximity of the local universe \citep[e.g.,][]{Koribalski2004,Meyer2004,Wong2006,Giovanelli2015,Haynes2011,Haynes2018}.

Over the past decade, a number of significant surveys have been conducted on the \HI line providing insights into the properties of the extragalactic population in the local universe \citep{Giovanelli2015}. The Arecibo Legacy Fast ALFA Survey (ALFALFA) was the most sensitive, large scale extragalactic \HI survey and the first of its kind \citep{Giovanelli2005, Haynes2018}. The ALFALFA survey utilized the seven-beam Arecibo $L$-band Feed Array (ALFA) to conduct a drift-scan survey, and covered an area of approximately 7000 deg$^2$ over a redshift range of $-2000<{\rm c}z<18000$\,\kms~and cataloged 31502 galaxies. Its beam size is $3\dotmin8\times3\dotmin3$, and the final spectral resolution is 10\,\kms~at 1420\,MHz \citep{Haynes2011,Haynes2018}. The ALFALFA survey has provided important measurements to determine the faint-end profile of the \HI mass function (HIMF) \citep[e.g.,][]{Martin2010,Jones2018}, the HIMF variation with the environment \citep[e.g., ][]{Moorman2014,Jones2016,Jones2018}, the H{\footnotesize I}-bearing population differing from optically selected objects \citep[][]{Huang2012a,Huang2012,Gavazzi2013}, and establishing metrics for the normal \HI content of galaxies \citep[e.g.,][]{Toribio2011,Odekon2016}. Another large extragalactic \HI survey before ALFALFA was the \HI Parkes All Sky Survey \citep[HIPASS;][]{Barnes2001}. It has also made a great census of the local extragalactic \HI galaxy distribution in a coverage of about 30000 deg$^2$, identifying more than 5000 galaxies, which are mainly included in the HIPASS Bright Galaxy Catalogue \citep[BGC;][]{Koribalski2004}, the southern HIPASS Catalogue \citep[HICAT;][]{Meyer2004}, and the Northern HIPASS Catalogue \citep[NHICAT;][]{Wong2006}, respectively.

However, to deal with a number of open questions a significant improvement over HIPASS and even ALFALFA is required. These surveys could not sample a cosmologically fair volume since their average depth was inadequate, and their angular and spectral resolution as well as sensitivity were limited. Consequently, earlier surveys on \HI have mainly focused on H{\,\footnotesize I}-rich galaxies that are local, massive, and situated at the upper end of the HIMF. For instance, these surveys did not detect sufficient \HI components from the population of dwarf galaxies. Typically, dwarf galaxies exhibit weak \HI signals with narrow \HI widths. As a result, higher sensitivity is required with smaller velocity bins. Furthermore, sensitivity limitations prevent the identification of the vast \HI morphology of bright local galaxies. This understanding is critical to discern their recent formation or merger history \citep[e.g.,][]{Zhou2023}.

The Widefield ASKAP L-band Legacy All-sky Blind Survey (WALLABY) is expected to survey 75\% of the sky in the 21-cm line of neutral atomic hydrogen gas at a resolution of 30$''$. It is projected to detect approximately half a million galaxies in the local universe \citep{Koribalski2020}. As the next-generation extragalactic survey in the 21-cm \HI line, WALLABY is expected to be the most powerful survey in observing the local universe. However, it will take years to complete the WALLABY survey.

The Five-hundred-meter Aperture Spherical radio Telescope (FAST) has the distinction of being designated as the most powerful single dish telescope. It is located at a geographic latitude of $25^\circ39'10\dotsec6$. Its observable maximum zenith angle is about $40^\circ$ \citep{Nan2011,Jiang2019,Jiang2020}. The \textbf{F}AST \textbf{A}ll \textbf{S}ky \textbf{H\,{\footnotesize{I}}} survey (FASHI) will cover the entire sky that can be detected by FAST telescope, ranging a declination of $-14\degree$ to $+66\degree$, and has a frequency range of $1.0-1.5$\,GHz, achieves a median detection sensitivity of $\sim$0.76\,$\mjyb$ at velocity resolution of 6.4\,\kms~\citep{Jiang2019,Jiang2020}. A pilot FASHI drift scan survey released and compared a total of 544 extragalactic \HI sources with the ALFALFA survey \citep{Kang2022}. The FASHI and ALFALFA surveys have consistent observational parameters, including flux and velocity. However, in sensitivity, efficiency, and resolution, the FASHI project outperforms the ALFALFA significantly. Furthermore, the FASHI survey has successfully mapped extragalactic \HI sources, such as M106 \citep{Zhu2021}, M51 \citep{Yuhy2023}, DDO\,168 \citep{Yunp2023}, M94 \citep{Zhou2023}, NGC\,4490/85 \citep{Liu2023}. This further demonstrates the significant scientific value of FASHI data for both statistical studies and targeted investigations.

Being a blind \HI survey, FASHI avoids bias towards high surface brightness galaxies, which are usually present in optical galaxy catalogs, and offers better angular and spectral resolution in comparison to any previous single telescope. FASHI's extensive coverage overlaps with several other prominent surveys, including the Sloan Digital Sky Survey \citep[SDSS;][]{Abazajian2009}, the Panoramic Survey Telescope and Rapid Response System \citep[Pan-STARRS;][]{Chambers2016}, the Two Micron All Sky Survey \citep[2MASS;][]{Skrutskie2006}, and the NRAO VLA Sky Survey \citep[NVSS;][]{Condon1998}. The catalog products generated by FASHI will be highly valuable for multiwavelength data mining, extending far beyond those currently engaged in the FASHI project itself. The program aims to provide legacy data products that have broad applications and maximize the potential for scientific research.

In this paper, we mainly present the FASHI project and its catalogued extragalactic \HI sources. Moreover, Section\,\ref{sec:survey} explains the scientific objectives, survey strategy, and observational setup of the FASHI project. Section\,\ref{sec:data_reduc} discusses data reduction, which includes baseline correction, radio frequency interference (RFI) mitigation, source extraction, and distance calibration. Section \ref{sec:catalog} describes the FASHI catalog that has been released. It outlines the approach used for source characterisation. It also provides an overview of the basic parameters of the detected sources and presents the methods used for optical counterpart identification. Section\,\ref{sec:discuss} comprises of the evaluation of the reliability and completeness of the FASHI sources. It also provides some caveats and comments for the users of the FASHI catalog. A summary is provided in Section \ref{sec:summ}. This paper adopts the Hubble constant value of H$_{0}$ = 75\,\kms Mpc$^{-1}$, $\Omega_{\rm M} = 0.3$, and $\Omega_{\rm \Lambda} = 0.7$. In addition, we adopt AB magnitude system \citep{Oke1983} throughout this paper.

\section{The FASHI project}\label{sec:survey}

\begin{figure*}
\centering
\includegraphics[width=0.99\textwidth, angle=0]{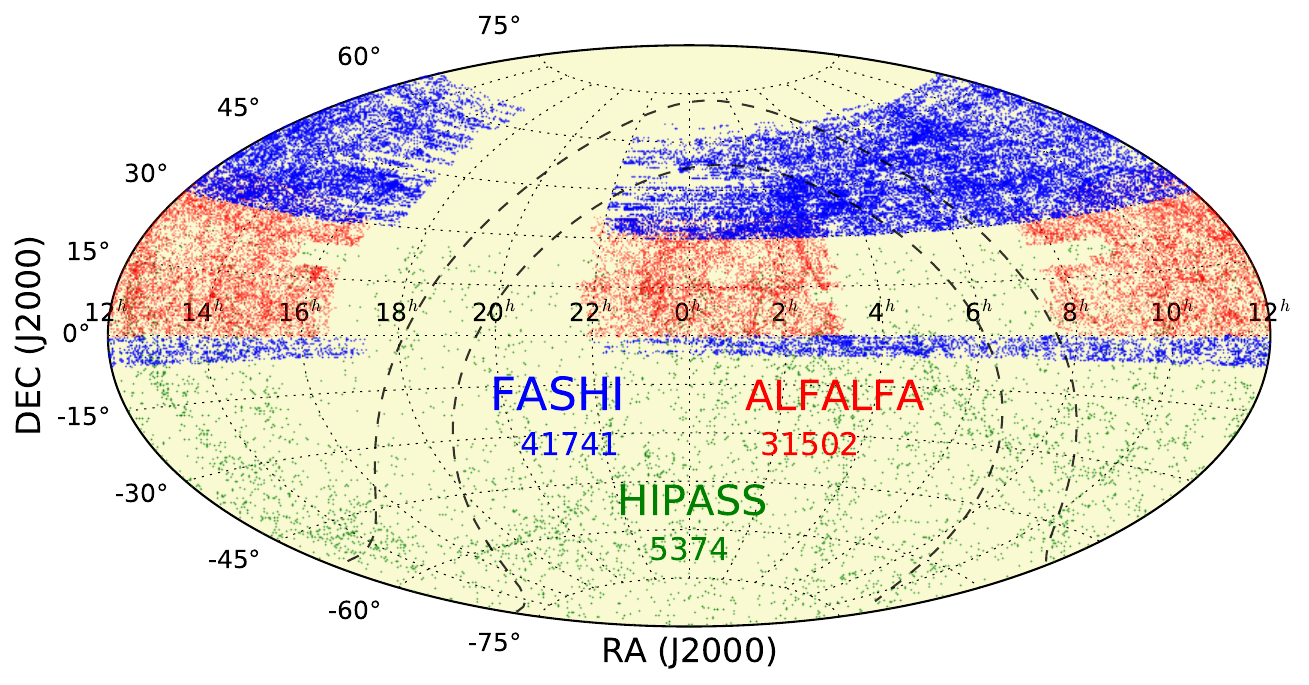}
\caption{FASHI sky distribution of currently released 41741 H{\,\scriptsize I} sources (in blue points) included in Table\,\ref{tab:exgalcat} in the Galactic hemispheres, showing the roughness of boundaries imposed by practical and scheduling constraints. For comparison, ALFALFA $\alpha$100 \citep{Haynes2018} and HIPASS galaxies \citep{Koribalski2004,Meyer2004,Wong2006} are also indicated with red and green points, respectively. The two black dashed lines show the position of the Galactic plane at galactic latitude $b=\pm10^\circ$.}
\label{Fig:observed_sky}
\end{figure*}

\begin{figure*}
\centering
\includegraphics[width=0.38\textwidth, angle=0]{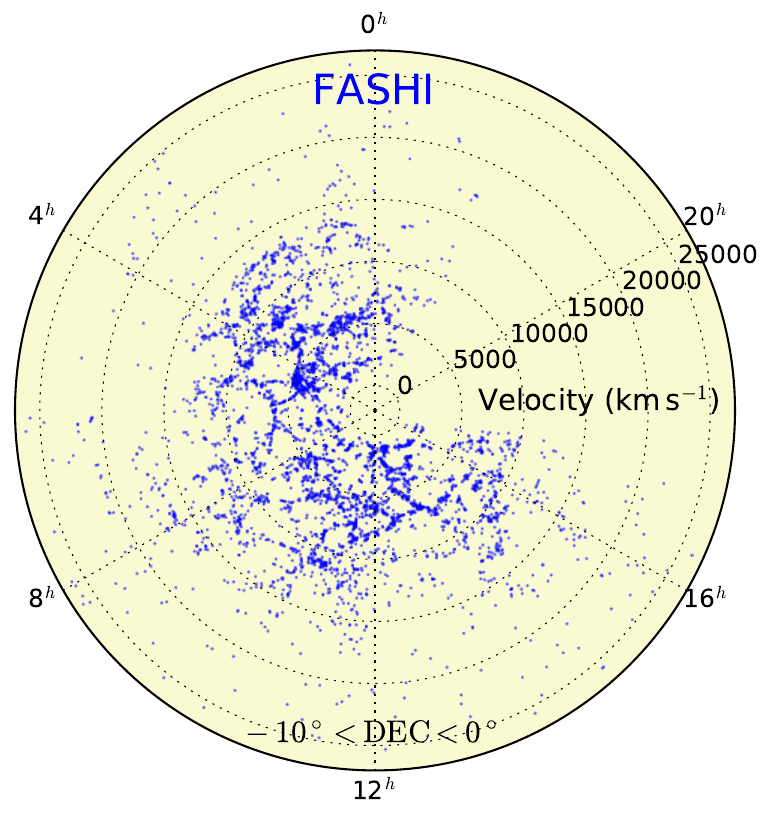}
\includegraphics[width=0.39\textwidth, angle=0]{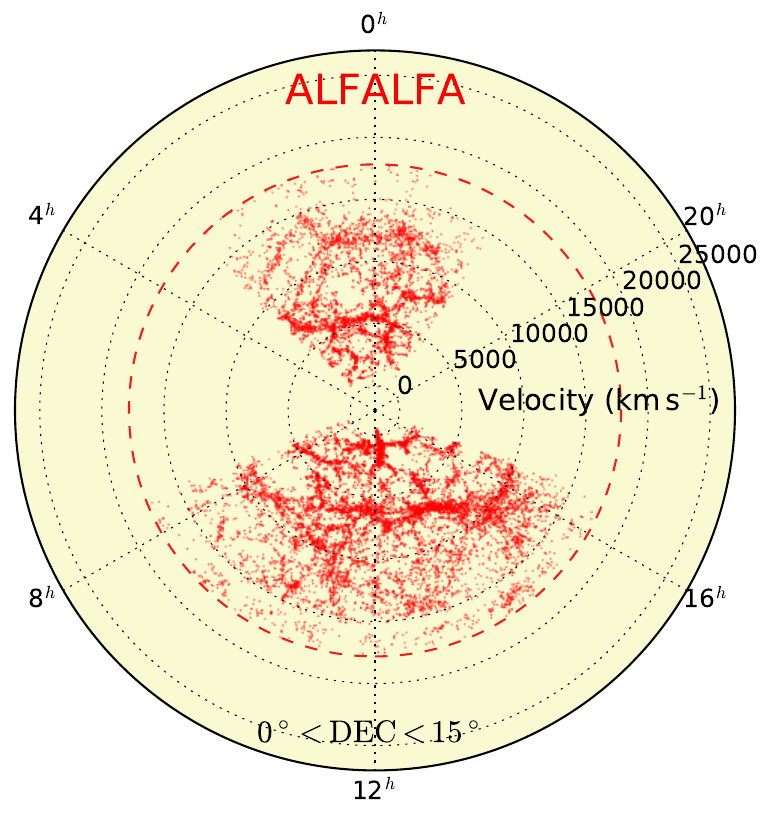}
\includegraphics[width=0.39\textwidth, angle=0]{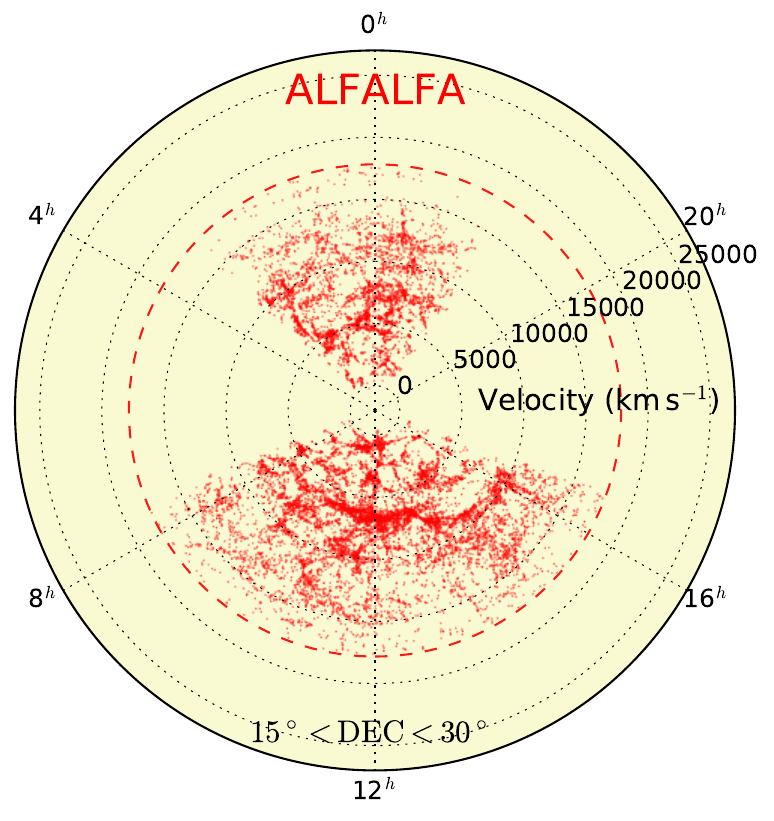}
\includegraphics[width=0.39\textwidth, angle=0]{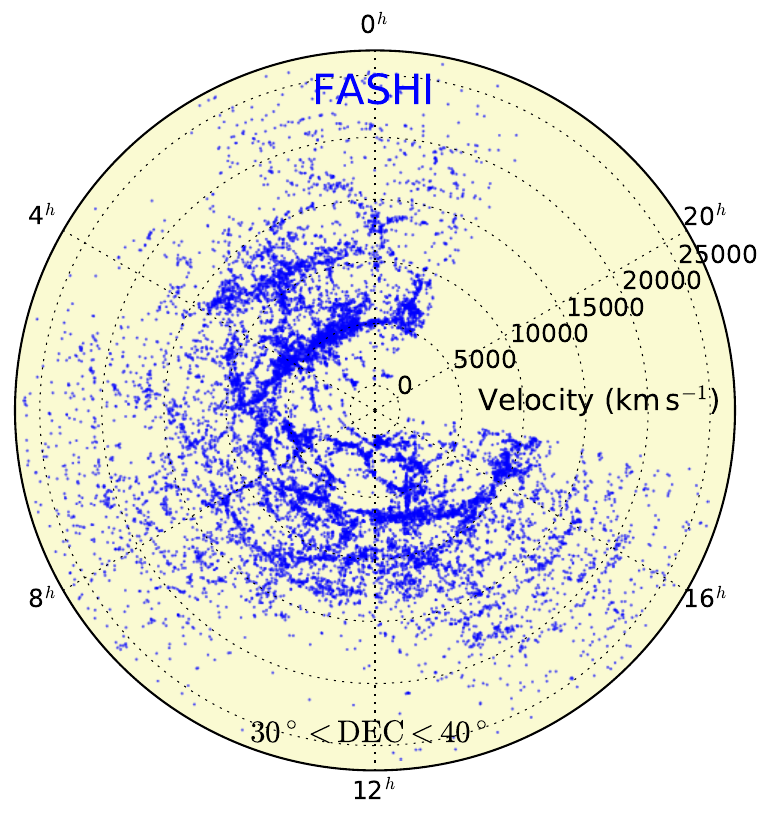}
\includegraphics[width=0.39\textwidth, angle=0]{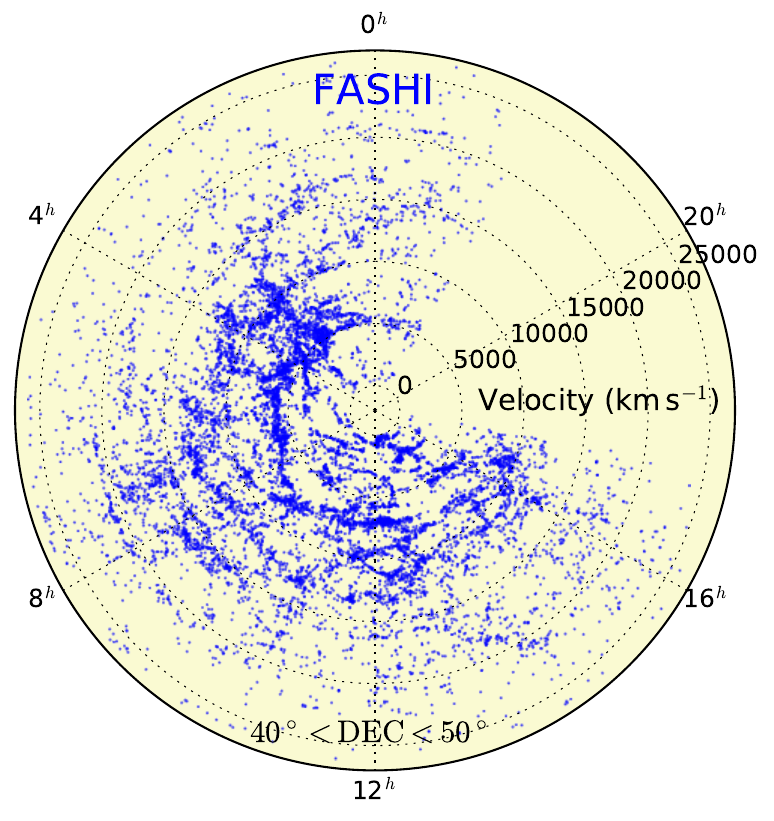}
\includegraphics[width=0.39\textwidth, angle=0]{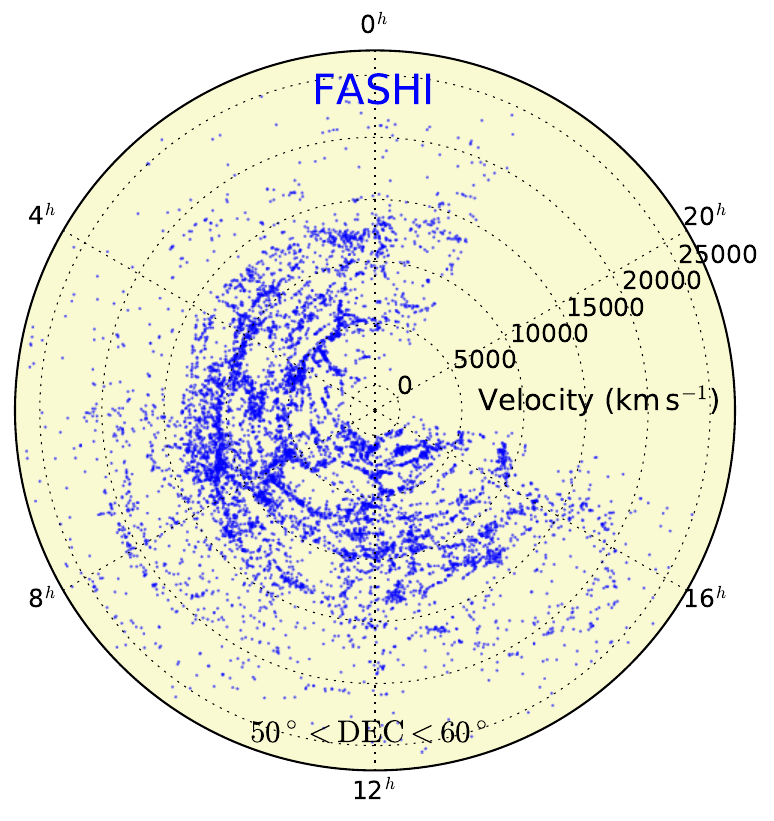}
\caption{The distribution of large-scale structure from FASHI (in blue points) and ALFALFA (in red points) sources at different declination (J2000) ranges of $\rm -10^\circ<DEC<60^\circ$ and volume extending to $\sim$26323\,\kms~that corresponds to the maximum velocity of current FASHI sources. The red dashed circle displays the maximum velocity of ALFALFA sources in the corresponding ALFALFA images.}
\label{Fig:polar}
\end{figure*}

The FASHI project aims to observe the entire detectable sky of the FAST in the declination range of $-14\degree$ to $+66\degree$ at a frequency between 1050 and 1450\,MHz, as illustrated in Figure\,\ref{Fig:observed_sky}. The first data release of the FASHI project mainly concentrates on extragalactic areas that lie predominantly outside the observable regions of the Arecibo telescope. Between August 2020 and June 2023, the FASHI project observed an approximate area of 7600 square degree of the sky and detected 41741 extragalactic \HI sources, as shown in Figure\,\ref{Fig:observed_sky}. This indicates a detection rate of approximately 5.5 sources per square degree. Based on this detection rate, FASHI is expected to detect over 100000 extragalactic \HI galaxies in the approximately 22000 square degree region detectable by FAST, up to a redshift of approximately 0.35. Currently, we mainly use datasets in the frequency range of 1305.5-1419.5\,MHz due to serious RFI and relatively low sensitivity at low frequency ranges. Table\,\ref{tab:obs} summarizes the technical details of the FASHI project.

\begin{table}[H]%\vspace{-0.45 cm}
\caption{\textbf{FASHI project technical details.}}
\label{tab:obs}
\vskip 5pt
\centering
\footnotesize{
\begin{tabular}{l|l}
\hline
\hline
%\startdata
Parameters                      & Values \\
\hline
Planed RA range                 & $\rm 0^h\leq RA \leq24^h$ (see Figure\,\ref{Fig:observed_sky}) \\
Planed DEC range                & $\rm -14\degree \lesssim DEC \lesssim +66\degree$ (see Figure\,\ref{Fig:observed_sky}) \\
Current RA range                & $\rm 0^h\leq RA \leq17.3^h$, $\rm 22^h\leq RA \leq24^h$  \\
Current DEC range               & $\rm -6\degree \lesssim DEC \lesssim 0\degree$, $\rm 30\degree \lesssim DEC \lesssim 66\degree$  \\
Receiver                        & 19-beam array \\
Polarizations per beam          & 2 linear (XX, YY) \\
Beam size (FWHM)                & 2$\dotmin$9 at 1420\,MHz \\
Gain                            & 13-17\,K\,Jy$^{-1}$ \\
$T_{\rm{sys}}$                  & 16-19 K \\
Total frequency coverage        & 1050-1450\,MHz \\
Released frequency range        & 1305.5-1419.5\,MHz (in this work) \\
c$z_{\odot}$ range              & 200-26323\,\kms\, (in this work) \\
Bandwidth (total)               & 400\,MHz \\
Spectral channels               & 65536 (before smoothing) \\
Channel spacing                 & 7.629\,kHz or 1.6\,\kms~at 1420\,MHz \\
Spectral resolution             & 6.4\,\kms (after smoothing) \\
Spectral median rms             & $\sim$1.50\,mJy at 6.4\,\kms resolution \\
Map median sensitivity         & $\sim$0.76\,$\mjyb$ at 6.4\,\kms resolution \\
%\enddata
%\end{deluxetable}
\hline
\hline
\end{tabular}}
\begin{flushleft}
\textbf{Notes.} The spectral median rms is estimated from the baseline measurement of each integrated spectrum. The map median sensitivity of the map is estimated from the integrated map. \\
\end{flushleft}
\end{table}

\subsection{Science goals}

The primary survey strategy of the FASHI project is to utilize the high sensitivity and rapid survey speed of FAST to perform a comprehensive inventory of the \HI gas in the neighbouring universe. Additionally, the project aims to observe at leas 100000 \HI galaxies.

The ALFALFA project identified 31502 extragalactic \HI sources, determined the HIMF of the local universe, established measurements for the typical \HI content of galaxies, and derived the clustering properties of \HI galaxies \citep[e.g.,][]{Haynes2011,Toribio2011,Odekon2016,Jones2018}. 
Nevertheless, the studies concerning the low-mass end of the HIMF from ALFALFA still face detection limitations. Increasing the \HI detection number by an order of magnitude towards low masses is necessary to investigate the faint end of the HIFM, particularly the properties of the \HI faint targets, which include dwarf galaxies and the passive galaxy population. According to \citet{Huang2012}, the population detected by \HI surveys presents an overall higher star formation rate at a fixed stellar mass. However, for low-mass star-forming galaxies (blue dwarf galaxies) and quiescent galaxies, the \HI detection rate is relatively low. According to a study by \citet{Ai2018} of \HI gas in galaxy groups and clusters using FASHI data, the detection rate of \HI is less than 30\% for member galaxies in groups with $z>0.03$. This low detection rate poses a significant challenge to statistical studies of \HI gas properties in optically selected samples. There is no doubt that deeper surveys, such as FASHI, can help to improve the detection rate for \HI deficient galaxies.

The FASHI project focuses on providing a more comprehensive and extensive coverage of the local universe, as visualized in Figure \ref{Fig:polar}. Additionally, it aims to achieve a survey volume that is comparable to current optical surveys such as the SDSS. The project would enable the study of the HIMF across various cosmic environments, making it possible to find more low-mass dwarf galaxies and map the local Voids accurately in order to sample multiple low-density environments. FASHI is sensitive to large-scale diffuse gas structure, which could be missed by interferometric surveys. One of the ultimate goals of the FASHI project is to combine the FAST data with the data from ongoing or planned surveys employing interferometric arrays, which will enable sampling of the extragalactic \HI sky with significantly increased resolution. This will provide crucial information regarding the large-scale \HI structures, leading to a better understanding of the role of \HI in galaxy formation and evolution. Therefore, single-dish large-scale surveys like FASHI are still indispensable even in the era of interferometric large-scale surveys.

It has been demonstrated by ALAFALA that a blind \HI survey has the potential to discover uncommon objects like the nearby faint dwarf Leo\,P \citep{Giovanelli2013}, nearly dark galaxies, and galaxies with extraordinarily high H{\,\footnotesize I}-to-stellar mass ratios \citep[e.g.,][]{Adams2015,Janowiecki2015,Janesh2015,Janesh2017}. In the local universe, deeper surveys of larger volumes may detect additional such objects and even new types of species. FAST's sensitivity could enable the identification of \HI clouds, close to $10^4\,\msun$, in the Local Group, within a few minutes of integration time. In several case studies \citep[e.g.,][]{Zhu2021,Yuhy2023,Yunp2023,Zhou2023,Liu2023}, FASHI has detected intergalactic and surrounding diffuse gas that has not been detected by other telescopes. The work of \citet{Xujl2023} shows that FASHI can potentially discover more candidates of ``dark galaxies" that do not form any star due to low-density gas and, therefore, reveal the mystery of dark halo substructure.

Deep \HI observations using mapping techniques have the potential to reveal filamentary gas features \citep[e.g.,][]{Zhu2021} and large-scale distributions (refer to Figure \ref{Fig:polar}) of matter in the Local Group and nearby universe. This can provide more insight into the distribution and properties of dark matter. The \HI gas that is faint and diffuse, and cannot be detected by an interferometer, can be easily searched using FAST. There will be no restriction on the survey's depth level, and the scope of exploration will be expanded in certain areas based on the discovery of significant characteristics.

In addition, the FASHI project will also cover a section of the Milky Way. This will provide an opportunity to detect numerous radio recombination lines and high-velocity Galactic \HI clouds with exceptional sensitivity.

\subsection{Survey strategy}
\label{sect:strategy}

Survey efforts of this scale and scope require careful optimization of their operational strategy towards achieving the science objectives within the constraints imposed by practical observing conditions and requirements. In this paper, the available observation time for the FASHI project mainly comes from the schedule-filler time, i.e., the observations were carried out as a time-filler project when there are no other programs in the FAST observing queue. Although such type of observations cannot be planned for specified targets, we can still do a blind search for \HI emission. Our observational setup was to fix the sky declination and do drift scans during the project time. We also used some commissioning time to do calibration observations. In this mode, we can not control the sky coverage uniformly, but it makes full use of all available time to increase the sample of \HI galaxies. Some areas will be sampled multiple times and can go deeper than other areas. Such strategy has the advantage of finding more dwarf galaxies and \HI poor galaxies. As this is the first data release of the survey, it is not a high priority for us to achieve a uniform sky coverage. Instead, we made full use of all available time to increase the coverage area. We plan to standardize the survey depth for this ongoing project and will showcase the results in the next data release.

\subsection{Observational setup}

FAST has a 19-beam receiver enabling efficient scan observations that can cover multiple areas of observation simultaneously. The FAST spectral-line backend is equipped with two linear polarizations, \texttt{XX} and \texttt{YY}. The spectral-line backend has a full wideband that covers a frequency range of 1050 to 1450\,MHz. The FASHI project used the low-resolution mode with 64\,k channels, resulting in a frequency resolution of 7630\,Hz. The sampling time for each spectral line in FASHI is 1\,second. The FAST has an aperture of 500\,m with an effective aperture of approximately 300\,m, resulting in an HPBW of 2$\dotmin$9 at a frequency of $\sim$1.4\,GHz. The pointing error is less than or equal to $15''$. To calibrate intensity, a high-noise signal with an amplitude of $\sim$11\,K was injected into the entire FASHI observation for a period of 32 seconds. The factor of degrees per flux unit (DPFU) is relevant for 19 beams, with a value of $\rm DPFU=13\sim17$\,K\,Jy$^{-1}$ at a frequency of 1400\,MHz, as detailed in \citep{Nan2011,Jiang2019,Jiang2020,Qian2020,Han2021}. The technical details for the FASHI measurements are listed in Table\,\ref{tab:obs}.

\subsubsection{\texttt{DecDriftWithAngle} mode}\label{sec:driftscan}

Although FAST has an active reflector that enables the telescope to cover a larger part of the sky, the motion of the reflector panels is driven by over 2000 actuators, which consume a substantial amount of electricity. The drift mode has been extensively used at the Arecibo telescope, and the ALFALFA survey has shown that it can deliver high-quality data with a very good baseline. Thus, the FASHI project is predominantly performed using the drift scan mode (\texttt{DecDriftWithAngle}).

The FASHI survey was mainly carried out using a drift scan observational mode called \texttt{DecDriftWithAngle}, utilizing 19-beam receivers \citep{Li2018,Jiang2019,Jiang2020}. The drift scan method helped minimize complexities in system control and self-generated RFI, ensuring relatively accurate pointing, full FAST gain, and stable status for the primary surface and focal cabin during the surveys. Throughout the survey, the telescope's azimuth arm was positioned on the meridian at a pre-assigned J2000 declination, with a spacing of $21\dotmin65$ between primary drift centers. The feed array was rotated by $23\dotdeg4$ to enable super-Nyquist sampling of Earth-rotation drift-scan tracks of individual beams in the declination of J2000 coordinates, with sampling rates smaller than half the FWHM.

\subsubsection{\texttt{MultiBeamOTF} mode}\label{sec:otf}

Among all the FAST approved projects, the FASHI project has the lowest priority of observation. For the FASHI project to proceed, it relies solely on the schedule-filler observable time. Please refer to section \ref{sect:strategy} for details. As a consequence of this, the observed RA ranges during the FASHI drift survey appear relatively random. While some sky ranges have been scanned repeatedly, some others cannot be covered at all. To cover the portions of the sky that remained uncovered in the drift-scan mode, the \texttt{MultiBeamOTF} scan mode was employed \citep{Jiang2020}. The observational setup used in the \texttt{MultiBeamOTF} mode in FASHI is almost identical to that in the \texttt{DecDriftWithAngle} mode (see Section \ref{sec:driftscan}), with the exception of a specified declination range.

\section{Data reduction}\label{sec:data_reduc}

The FASHI data were reduced mainly using the FAST spectral data reduction pipeline named as \texttt{HiFAST} \citep{Jing2023,Xu2023}. The \texttt{HiFAST} pipeline combines several useful data reduction packages including antenna temperature correction, baseline correction, RFI mitigation, standing-waves correction, gridding, flux correction, and fits cube generation. For the FASHI data, the spectral resolution was smoothed to 6.4\,\kms~per channel, and the fits cubes were gridded into a pixel scale of 1\arcmin, meeting the Nyquist sample criterion. The conversion from antenna temperature to flux density was carried out using the fitted gain curves for FAST 19-beams when gain varied with different zenith angles ($\rm ZA\lesssim40\degree$), as reported by \citet{Jiang2019}. All other parameters used for data reduction were set to default values as specified in the \texttt{HiFAST} pipeline cookbook\footnote{\url{https://hifast.readthedocs.io/}}.

\subsection{The \texttt{arPLS} for baseline correction}\label{sec:arpls}

In the \texttt{HiFAST} pipeline \citep{Jing2023}, we utilized the Asymmetrically Reweighted Penalized Least Squares algorithm \citep[\texttt{arPLS};][]{Baek2015} to adjust the entire 500\,MHz spectral baseline. This method permits us to modify a parameter, ``weight'', repeatedly to calculate the baseline. A significant weight is applied automatically when the signal falls below an already fitted baseline. Conversely, no weight or a small weight is assigned when an emission line is present above the fitted baseline as it may be part of the spectral peak. A preset coefficient can control the balance between fitness and smoothness, based on the total spectral channel number. This method can iteratively estimate the noise level and adjust the weights accordingly \citep{Baek2015}. Thankfully, the FASHI data contains only a few instances of weak and narrow absorption dips throughout the entire 500\,MHz band. Therefore, it is highly suitable to implement the \texttt{arPLS} method for reducing the FAST spectral data. In fact, the \texttt{arPLS} baseline fitting method has been effectively employed for many FAST spectral analyses \citep[e.g.,][]{Zeng2021,Zhang2021,Zhang2022}. Nevertheless, there is a drawback in the arPLS algorithm for baseline correction. Our findings revealed that a small indentation occasionally appears in the baseline close to the line wings of the strong emission and wide line profile with intensities over $\sim$40\,mJy and velocities over $\sim$150\,\kms. This could lead to a slight underestimation ($\lesssim$5\%) of the integrated flux only for bright sources. In general, \texttt{arPLS} is the best choice for baseline correction when reducing broadband data from large scale surveys.

\subsection{Radio frequency interference}\label{sec:rfi}

The quality of antenna observational data is seriously affected by RFI. Mitigating RFI is an essential procedure in any observation. However, strong RFI from sources such as broadcast radio and TV, communication satellites, and navigation satellites are extremely difficult to avoid in FASHI. As a result, the existing RFI have made several wide frequency bands completely useless. \citet{Zhang2022} conducted a statistical analysis of the probabilities of RFI distribution, derived from approximately 300 hours of blind survey data, across the wideband FAST spectrum from 1050 to 1450\,MHz. RFI is relatively severe below 1305\,MHz, whereas the frequency band above 1305\,MHz is relatively clear. At approximately 1382\,MHz, a strong RFI is present due to the GPS L3 signal. This corresponds to a distance of around 112\,Mpc as shown in Figure \ref{Fig:mass_distance}. Thankfully, the probability of the strong RFI appearing is less than 10\%. These occasional RFIs have been removed reasonably well during the data reduction process. We flag RFIs in FAST spectra employing four strategies - flagging extremely strong RFIs, long-lasting RFIs, polarized RFIs, and beam-combined RFIs, as detailed in \citet{Zhang2022}. Therefore, we publish only the extragalactic sources that are located within the frequency band of 1305.5-1419.5\,MHz in the first released FASHI data.

Several frequency periods ($\sim$1\,MHz, $\sim$2\,MHz, and $\sim$0.04\,MHz) of standing waves are inhabited in the FASHI spectral data. The spectral quality can be significantly improved if the standing wave is completely corrected. Standing waves are predominantly fitted and subtracted using the SW package \citep{Xu2023} in the \texttt{HiFAST} pipeline \citep{Jing2023}. To find standing waves at the frequency periods in the FAST spectra, the SW package uses the Fourier transform method. To prevent confusion from the Galactic \HI emission, the velocity range of $-150<v_{\rm HI}<150$\,\kms~was masked and excluded from the spectral data reduction, as a further measure. The effect obtained after implementing the SW package correction shows that the standing wave is almost disappear. This treatment turns out to be highly effective for both standing wave correction and source extraction.

\subsection{Source extraction}\label{sec:extraction}

The process of extracting sources for FASHI consists of two steps, involving both automated and interactive procedures. The application used for finding sources in \HI data cubes is called the \HI Source Finding Application\footnote{\url{https://github.com/SoFiA-Admin/SoFiA-2}} (\texttt{SoFiA}) in version 2. \texttt{SoFiA} is a pipeline that was initially intended for detecting and characterising galaxies in extragalactic \HI data cubes in a three-dimensional space \citep{Serra2015,Westmeier2021,Westmeier2022}. The \texttt{SoFiA} source finding algorithm involves smoothing the data on various spatial and spectral scales specified by the user and measuring the noise level in each smoothing iteration. Subsequently, a user-defined flux threshold is applied relative to the noise to retain all pixels with statistically significant flux density. \texttt{SoFiA} is a fully automated and reliable source finding tool, making it crucial for the scientific success of surveys \citep[e.g.,][]{Westmeier2022}.

In the FASHI source search, we first used version 2 of \texttt{SoFiA} to generate a list of initial candidates. In the \texttt{SoFiA} setup, the detection threshold is 4.5$\sigma$ ($\sigma$ means the local noise level within the bounding box of the source in native units of the data cube; see the \texttt{SoFiA} cookbook); the smoothing kernels are \texttt{kernelsXY = 0, 3, 6} and \texttt{kernelsZ = 0, 3, 7, 15} respectively; in the linking parameters, the minimum size of the sources in the spatial dimension is 5 pixels/channel each in XY and Z space, while the maximum size is 50 pixels in XY space, but not limited in Z space; furthermore, we set the \texttt{reliability.enable} as false so that artefacts such as RFI and bad baselines occasionally remain in the spectral data. Based on the above setup, we found that the the source candidates found by \texttt{SoFiA} found contain some false signals. Therefore, a manual interactive source extraction has to be done, judging by the source moment maps, spectral profile, signal to noise ratio and so on. In addition, in the cases of double or multiple sources very close in position and redshift and low intensity level between two peaks, \texttt{SoFiA} has difficulty distinguishing them and often prefers to fit them with a wide double horn profile rather than two different single line profiles. For the cases of bad baselines produced by the \texttt{HiFAST} pipeline, we used a high-order polynomial fit to correct each source baseline again. In this step (see also Figure\,\ref{Fig:fast_sample}) we extend the integrated area to $\sim$2.0-2.5 times (see analysis in Section\,\ref{sec:catalog_col}) the 4.5$\sigma$ region size (see above for the definition of 4.5$\sigma$) for measuring each source integrated spectrum to include most of the flux from \HI sources. The size of the integrated area also depends on the SNR of each \HI line. If \texttt{SoFiA} did not fit the source position and size well, we would manually use an ellipse or circle of an appropriate size to measure the integrated spectrum of the source. The ellipse or circle size would try to include $\sim$100\% of each source's integrated flux, but a high enough SNR ($\rm SNR\gtrsim6.5$) for each source should be guaranteed first. Finally, we use the busy function to fit the integrated spectra. The velocity, line width and flux density would be derived by the busy function fitting code \citep{Westmeier2014}. Although all detections made by \texttt{SoFiA} were visually inspected to discard obvious false positives caused by noise peaks or artefacts, a small number of false detections are likely to remain in the final catalog. 

\subsection{Source aperture}\label{sec:size}

A source aperture, defined as an ellipse ($ell_{\rm maj,\,min,\,pa}$) with major and minor radii, and a position angle, is used for measuring flux. The entire \HI flux of each source is integrated within the ellipse, providing an estimate of the size of each \HI source. Usually, the size of the ellipse is set to be approximately 2.0-2.5 times larger than the area covered by the 4.5$\sigma$ region (refer to Section\,\ref{sec:extraction} for the definition of 4.5$\sigma$). Approximately 95\% of the measured effective diameter ($2\sqrt{ell_{\rm maj} \times ell_{\rm min}}$) in our catalog is larger than 5$\dotmin$0, with a median value of 8$\dotmin$3 which indicates that the current ellipse size is large enough compared to the beam size ($\sim$2$\dotmin$9) to encompass most of the emission from each \HI source. Furthermore, the size of the integrated area is dependent on the signal-to-noise ratio (SNR) of each \HI line. The catalog indicates that 94\% of sources have an SNR of at least 10, ensuring sufficient SNR for most sources. Reducing the size of the ellipse would lead to the loss of flux, whereas increasing it would introduce more noise. Our statistical analysis shows that changing the aperture of each source in the catalog by 10\% results in a decrease or increase in the included source flux of only $\sim$2.5\% or 2.0\% respectively. Similarly, the flux uncertainty reduces by roughly 11.5\% or 13.0\%, respectively. Furthermore, the current source aperture fulfils the criterion that the flux uncertainty is less than 10\% of each source flux. An increase of the source aperture by 10\% will elevate the contribution of flux uncertainty to roughly 12\% of the total flux of each source. Overall, the current setup of the ellipse size proves to be a highly suitable solution.

\subsection{Distance calculation}\label{sec:distances}

\begin{figure}[H]
\centering
\includegraphics[width=0.48\textwidth, angle=0]{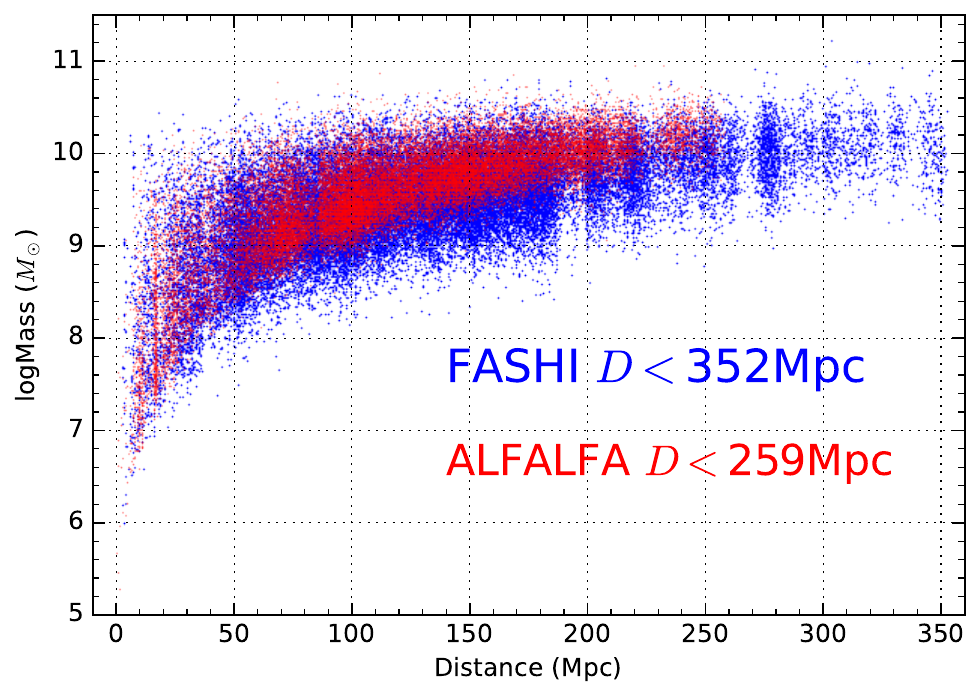}
\caption{FASHI mass-distance diagram of currently released 41741 H{\,\scriptsize I} sources (in blue points) listed in Table\,\ref{tab:exgalcat}. The strange gaps of the FASHI targets at $\sim$190, $\sim$210 and $\sim$266\,Mpc are probably caused by relatively low sample completeness. For comparison, the ALFALFA $\alpha$100 sources \citep{Haynes2018} are also shown here with red dots. The gaps of the ALFALFA targets at $\sim$80 and $\sim$230\,Mpc are caused by the known RFIs.}
\label{Fig:mass_distance}
\end{figure}

The peculiar motions of galaxies interfere with distance calibration in the local universe because of their gravitational interaction with the underlying density field of matter. This can be observed in the large-scale structures (e.g., Figure \ref{Fig:polar}) of the local universe \citep{Branchini1999, Masters2005, Masters2006}. Peculiar velocities of galaxies provide a powerful and unbiased method to comprehend the dynamics and structure of the local universe. Cosmicflows\footnote{\url{https://edd.ifa.hawaii.edu/}} is a program that compiles galaxy distances and breaks down observed velocities into components due to the expansion of the universe and the residual effects of gravitational interactions \citep{Graziani2019,Kourkchi2020}. Galaxy test particle peculiar motions can be used to derive inferences about the large-scale structure of the universe \citep{Tully2008, Tully2013, Tully2016}. The Cosmicflows program derives two Distance-Velocity Calculators that establish the connections between the distances and velocities of galaxies, based on smoothed versions of the velocity fields, with the aid of the Extragalactic Distance Database \citep{Tully2009}. One of this calculator is limited to a distance of 38\,Mpc and is based on a fully nonlinear analysis. The other calculation extends to a distance of 200\,Mpc and is derived from an analysis in the linear dynamical region \citep{Graziani2019,Kourkchi2020}. The distance measurement of the sources in FASHI mainly combines two distance calculators: the Cosmicflows-3 Distance-Velocity Calculators and the Hubble flow distance, with H$_{0}$ = 75\,\kms Mpc$^{-1}$. Details about the combination for the distance calculators are presented in Section\,\ref{sec:catalog_col}. The derived distance parameters are presented in Table\,\ref{tab:exgalcat} and illustrated in Figure\,\ref{Fig:mass_distance}.

\section{The FASHI extragalactic source catalog}
\label{sec:catalog}

\subsection{Extragalactic \HI catalog}
\label{sec:catalog_col}

\begin{table*}
\caption{\textbf{FASHI extragalactic H{\scriptsize{\,I}} source catalog.}}
\label{tab:exgalcat}
\vskip 5pt
\centering \small  %\footnotesize %\scriptsize
\setlength{\tabcolsep}{1.3mm}{
\begin{tabular}{cccccccccccccccc}
\hline \hline
[1]  &  [2]  & [3]   & [4]  & [5] & [6]  & [7]  & [8] & [9] & [10] & [11] & [12] & [13] & [14]  \\ 
FASHI ID &  RA & DEC &   c$z_{\odot}$ &  ell$_{\rm maj,\,min,\,pa}$ & W$_{50}$ & W$_{20}$ & $F_{\rm peak} $& $\sigma_{\rm rms}$  & $S_{\rm bf}$  & $S_{\rm sum}$  & SNR & $D$ & log$M$  \\
 & deg & deg   & \kms    & $~~',~~',~~^\circ$   & \kms     & \kms     & mJy & mJy& mJy$\cdot$\kms & mJy$\cdot$\kms &  & Mpc & $M_{\odot}$       \\
\hline
20230000823&166.736&-6.193&6802.1&3.7,\,3.7,\,0&164.0&172.6&9.0&1.8&1179.4&1046.8&20.3&98.2&9.4\\ 
20230000824&113.607&-6.193&10797.4&4.4,\,4.4,\,0&149.7&199.5&16.5&2.9&2221.2&2150.5&25.0&143.7&10.0\\ 
20230000825&166.377&-6.187&7596.6&5.0,\,4.9,\,106&33.0&46.5&33.2&2.4&1118.9&1098.3&31.9&108.3&9.5\\ 
20230042078&169.097&-6.182&24776.7&5.0,\,5.0,\,0&94.5&120.8&7.6&2.8&648.5&568.0&9.4&327.6&10.2\\ 
20230000826&111.686&-6.182&19246.1&2.3,\,2.3,\,0&198.6&216.4&2.4&1.2&474.5&504.0&11.2&253.7&9.8\\ 
20230042079&165.515&-6.178&9428.8&6.0,\,3.8,\,174&249.8&284.5&7.1&2.7&1403.0&1201.0&13.2&134.6&9.8\\ 
20230000828&163.988&-6.174&10202.9&5.8,\,5.4,\,33&170.9&176.9&23.4&3.4&2407.1&2550.6&21.2&143.9&10.1\\ 
20230000831&58.707&-6.171&10984.3&3.3,\,3.3,\,0&133.7&231.8&7.8&2.5&1022.5&984.2&13.7&152.0&9.7\\ 
20230042080&125.705&-6.169&14644.7&5.5,\,4.5,\,142&70.7&101.2&13.9&3.6&994.4&715.8&13.1&196.1&9.9\\ 
20230000829&110.411&-6.163&1641.4&6.7,\,6.5,\,109&183.1&258.9&16.0&4.2&3078.2&3119.5&21.4&24.5&8.6\\    
... ... \\
\hline
\end{tabular}}
\begin{flushleft}
\textbf{Notes.} The complete catalog containing 41741 sources can be accessed in the online Supplementary material or at \url{https://fast.bao.ac.cn/cms/article/271/} and \url{https://zcp521.github.io/fashi}, with additional columns available for download. \\
\end{flushleft}
\end{table*}

\begin{figure*}
\centering
\includegraphics[width=0.45\textwidth, angle=0]{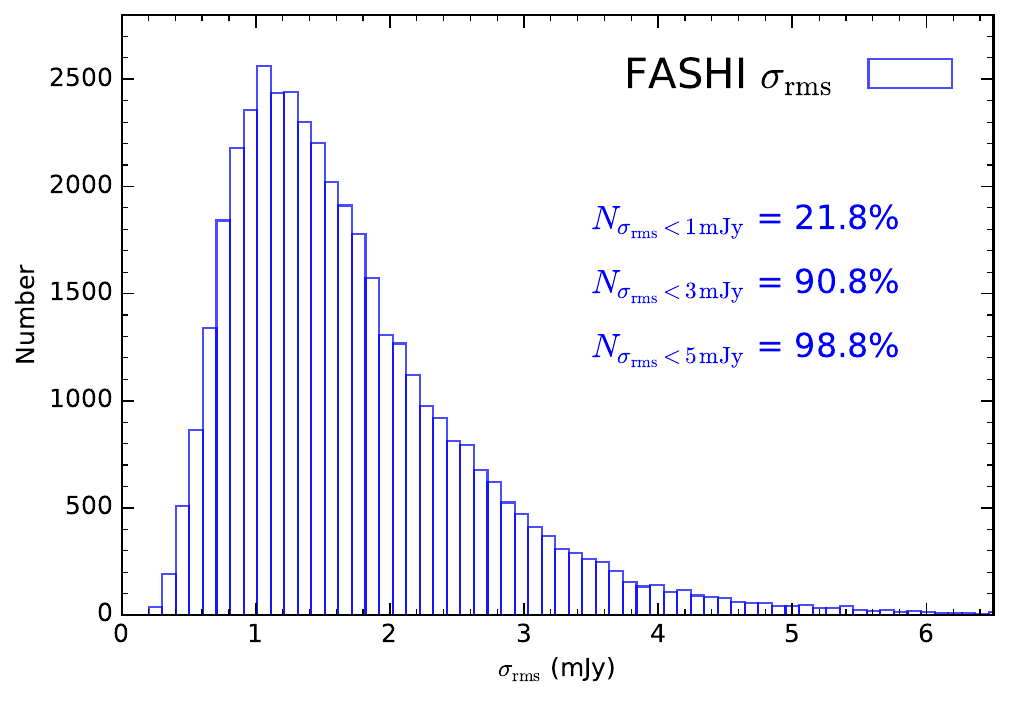}
\includegraphics[width=0.45\textwidth, angle=0]{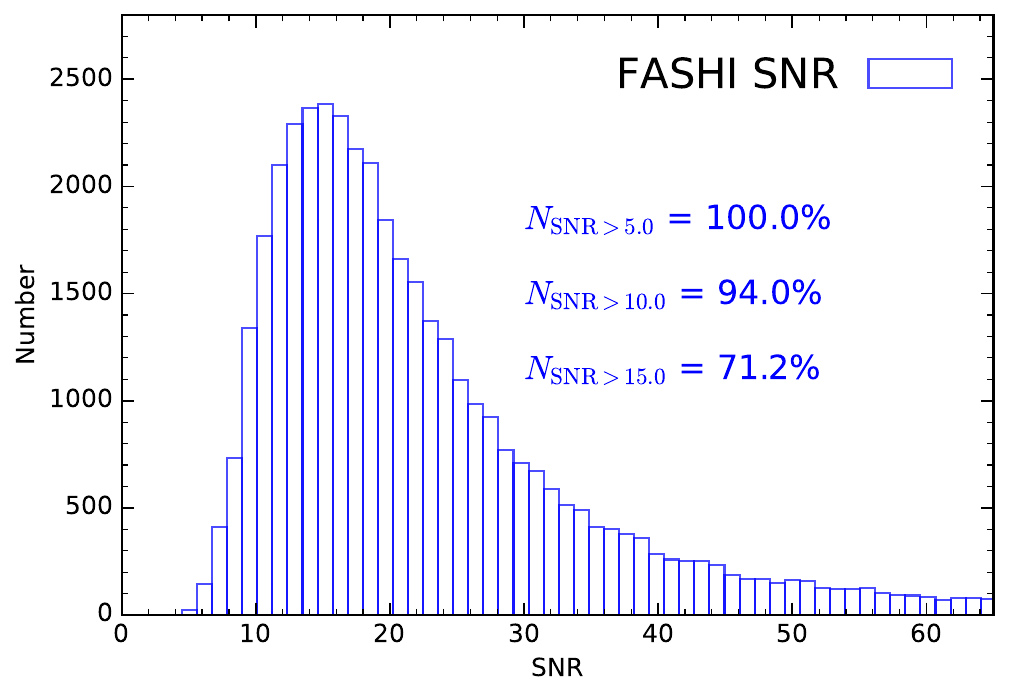}
\includegraphics[width=0.45\textwidth, angle=0]{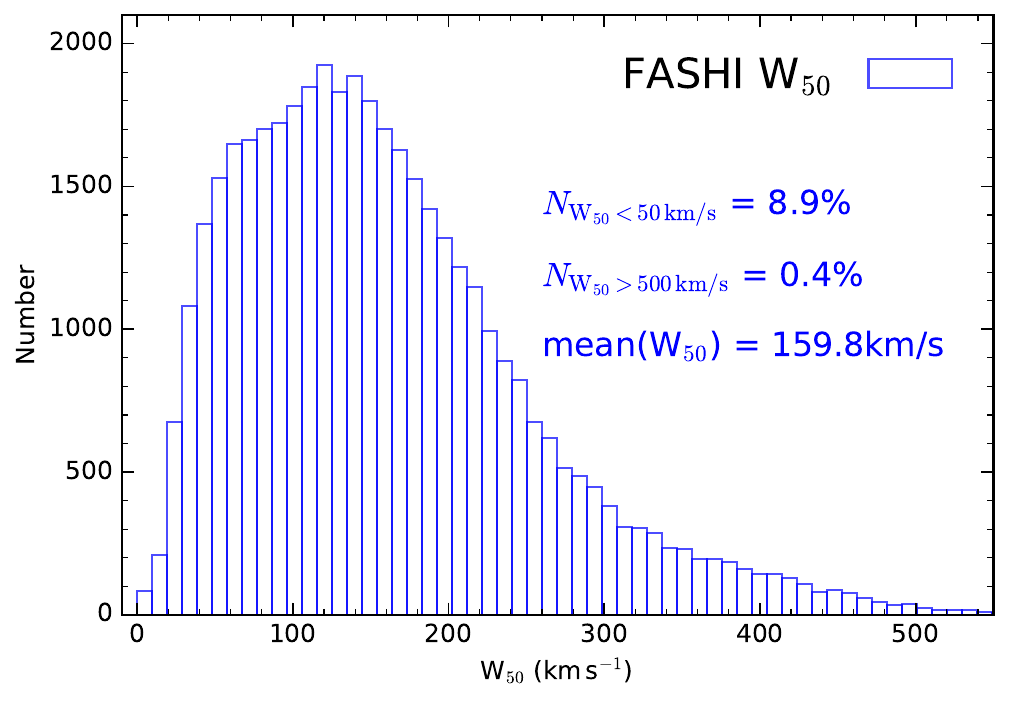}
\includegraphics[width=0.45\textwidth, angle=0]{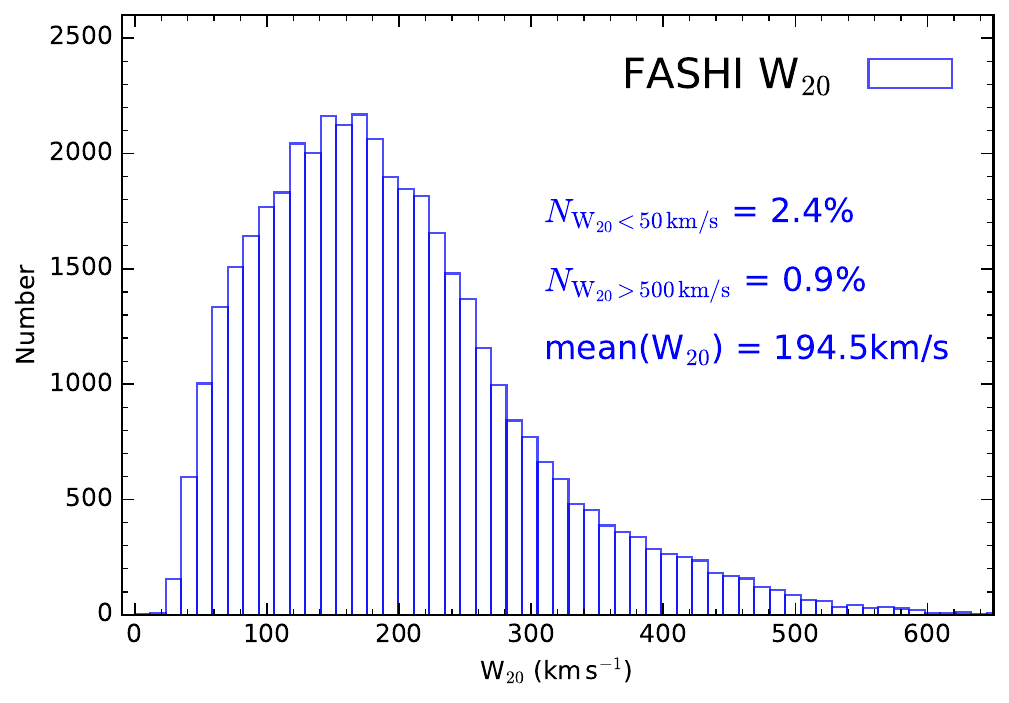}
\includegraphics[width=0.45\textwidth, angle=0]{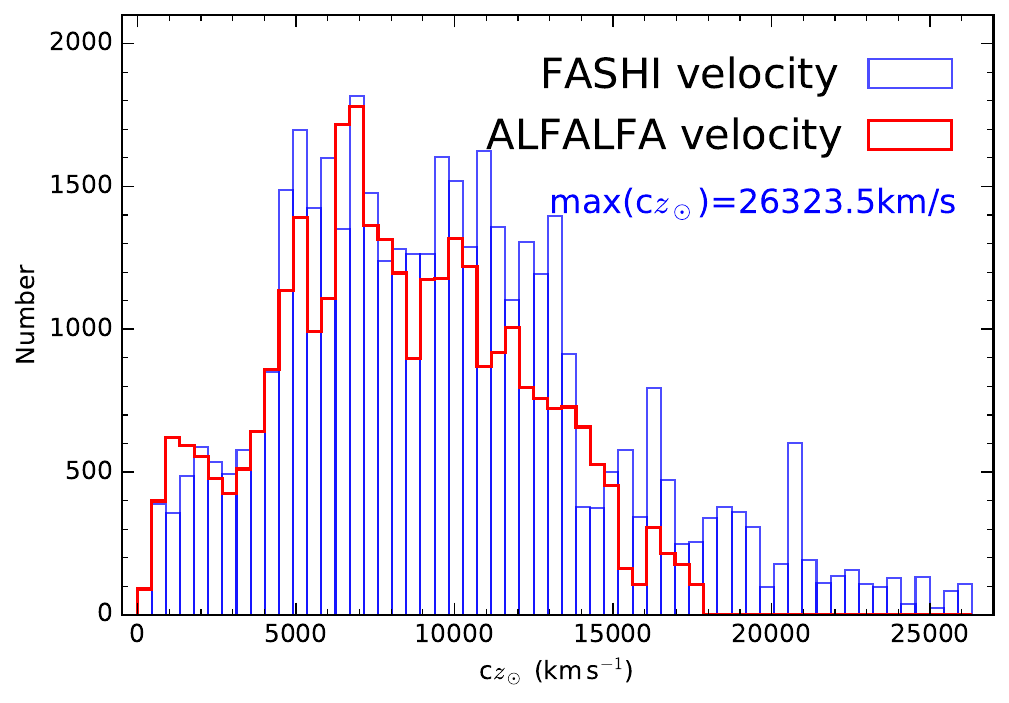}
\includegraphics[width=0.45\textwidth, angle=0]{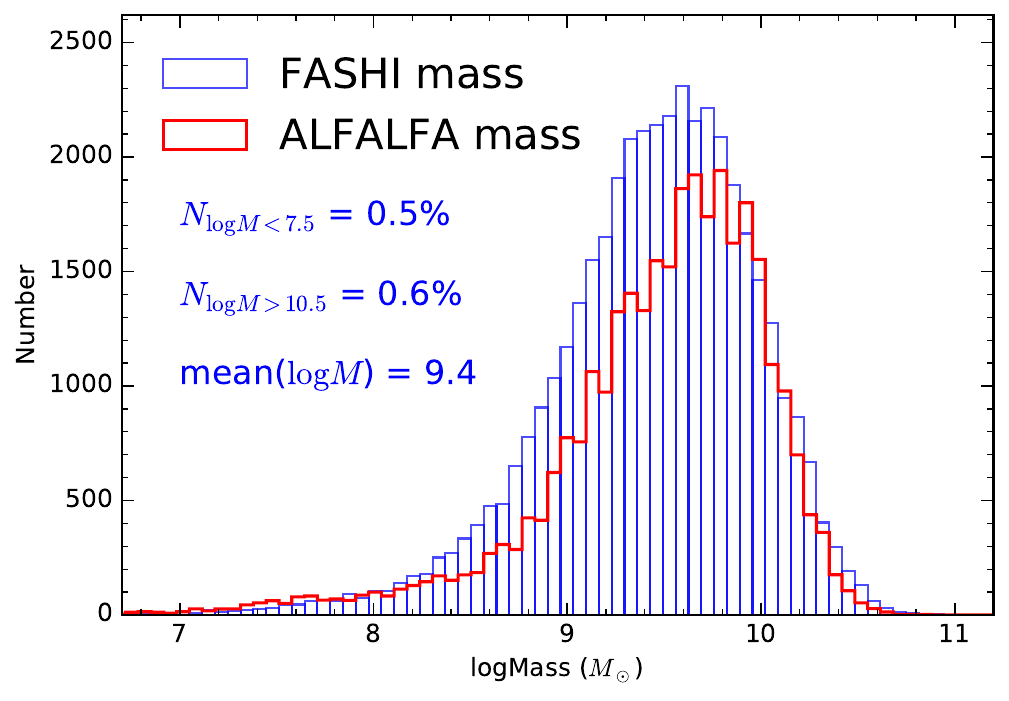}
\caption{Histograms of the measured spectral rms ($\sigma_{\rm rms}$), signal-to-noise ratio (SNR), line width $W_{50}$ and $W_{20}$, heliocentric velocity (c$z_{\odot}$) and source mass distributions for the entire released FASHI and ALFALFA catalogs. Some useful statistical results of the FASHI sources are given in each panel. For comparison, the source mass measured by ALFALFA has been roughly corrected with the same Hubble constant as that used by FASHI.} 
\label{Fig:hist_fast}
\end{figure*}

\begin{figure*}
\centering
\includegraphics[width=0.48\textwidth, angle=0]{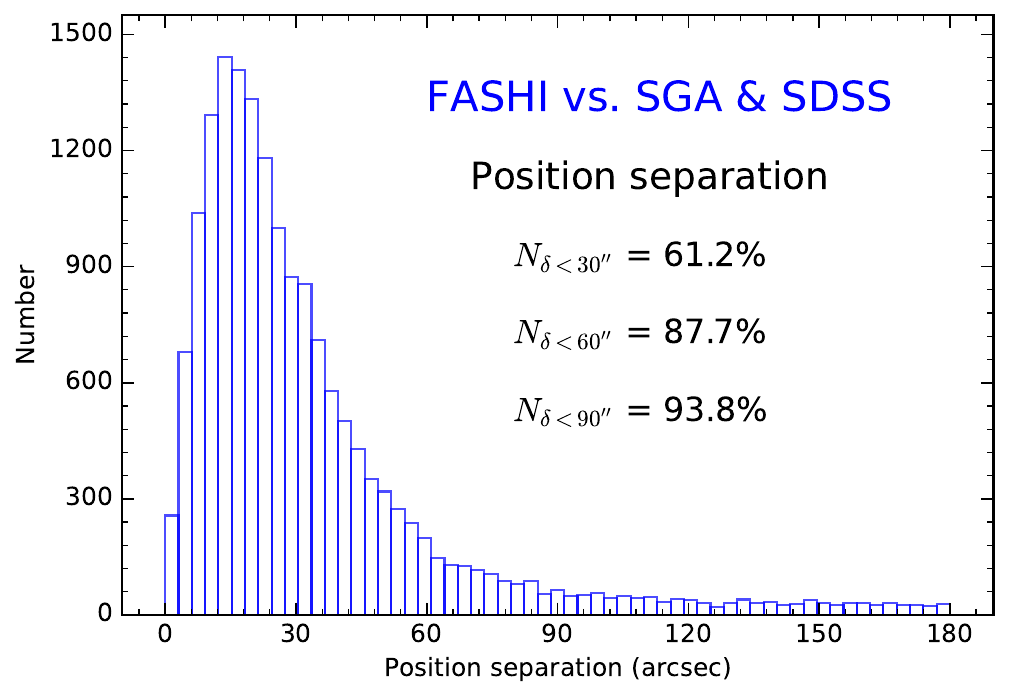}
\includegraphics[width=0.48\textwidth, angle=0]{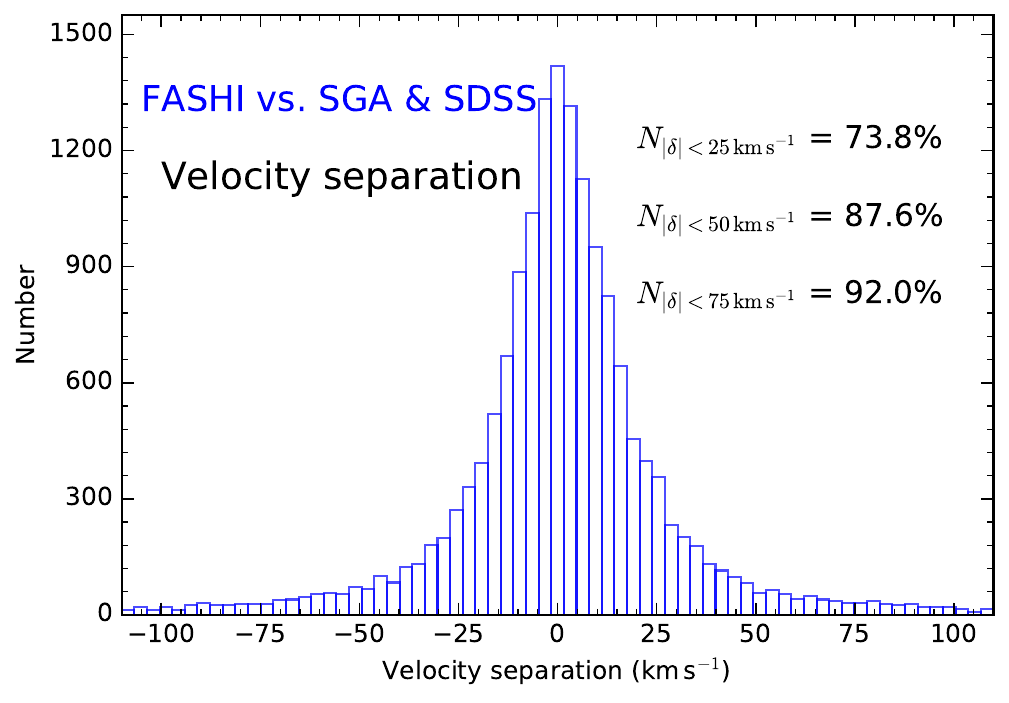}
\caption{The position and velocity separations between the FASHI and SGA \& SDSS spectroscopic sources obtained from the cross-matched samples (16972 sources) under a condition of $\rm\delta_{RA} \leq 3'$, $\rm\delta_{DEC} \leq 3'$ and $\rm\delta_{velocity} \leq 1000$\,\kms~listed in Tables\,\ref{tab:cross_sga} and \ref{tab:cross_sdss}. The occupied percentage within some separation ranges is also shown in each panel.}
\label{Fig:hist_match_sdss}
\end{figure*}

Table \ref{tab:exgalcat} presents the key findings of the FASHI extragalactic \HI survey. The complete contents of Table \ref{tab:exgalcat} can be downloaded in the online Supplementary material or at \url{https://fast.bao.ac.cn/cms/article/271/} and \url{https://zcp521.github.io/fashi}. It comprises of 41741 sources having multiple columns of observed and calculated parameters. The full online table provides high-precision and detailed information such as the centroid coordinate (J2000) in format of \texttt{Jhhmmss.ss$\pm$ddmmss.s}, spectral line central frequency in MHz, radio-defined velocity, integrated velocity range, detection sensitivity, and uncertainty. The introduction of each column in Table \ref{tab:exgalcat} and the parameters that are only included in the online table, are all introduced in the relevant column as follows:

\begin{itemize} 

\item {Column\,1: Index number for each FASHI extragalactic source. This index number is unique to each FASHI source.}

\item {Column\,2-3: Right ascension (RA) and declination (DEC) in unit deg of FASHI source centroid (J2000).  The distribution of the spatial flux of the sources often displays an irregular structure, or RFI and nearby objects can influence it, causing the source position offset to exceed the telescope pointing accuracy of approximately 7.9$''$ \citep{Jiang2020}. However, it should generally be less than 1$'$, as evidenced by the position separation between FASHI and SDSS sources in Figure\,\ref{Fig:hist_match_sdss}. For each FASHI source, the positional accuracy listed in the online table can be estimated by dividing the beam size ($\sim$2$\dotmin$9) by the \HI signal-to-noise ratio \citep[e.g.,][]{Koribalski2004}. In addition, the centroid coordinate (J2000) is only listed in the online table, formatted as \texttt{Jhhmmss.ss$\pm$ddmmss.s}. }

\item {Column\,4: Heliocentric velocity of each \HI source, c$z_{\odot}$ in unit of \kms. Values have been converted to the optical definition $\delta\lambda/\lambda$ from the ``radio'' one ($\delta\nu/\nu$). The c$z_{\odot}$ histogram of all FASHI sources is shown in Figure\,\ref{Fig:hist_fast}, where the strange gaps of the FASHI targets at $\sim$14300\,\kms\, are probably caused by relatively low sample completeness. The velocity uncertainty is estimated with $\sigma({\rm c}z_{\odot}) = 3\sqrt{P\,\delta v}/{\rm SNR}$, where $\delta v=6. 4$\,\kms~is the spectral resolution, and $P = (W_{20}-W_{50})/2$ is a measure of the steepness of the profile edges \citep{Fouque1990,Koribalski2004}. The spectral line centre frequency, radio velocity, redshift, $z_{\odot}$ and their uncertainties, corresponding to the heliocentric velocity, are only given in the online table. }

\item {Column\,5: Ellipse, $ell_{\rm maj,\,min,\,pa}$, with major, minor radius and position angle, in units of arcmin, arcmin, and degree, respectively. The ellipse indicates the measurement aperture of the integrated source in RA and DEC space for estimating the source flux. It does not represent the properties of the galaxy.} 

\item {Column\,6: Velocity width of the \HI line profile, $W_{50}$ in \kms, measured at 50\% level of every peak by busy-function fitting. The $W_{50}$ uncertainty is estimated with $\sigma(W_{50})=2\sigma({\rm c}z_{\odot})$. The $W_{50}$ histogram of FASHI sources is shown in Figure\,\ref{Fig:hist_fast}. }

\item {Column\,7: Velocity width of the \HI line profile, $W_{20}$ in \kms, measured at 20\% level of every peak by busy-function fitting. The $W_{20}$ uncertainty is estimated with $\sigma(W_{20})=3\sigma({\rm c}z_{\odot})$. The $W_{20}$ histogram of FASHI sources is displayed in Figure\,\ref{Fig:hist_fast}.}

\item {Column\,8: \HI line flux intensity at the peak of each integrated spectrum, $F_{\rm peak}$ in mJy, measured by busy-function fitting. The $F_{\rm peak}$ uncertainty is estimated with the measured $\sigma_{\rm{rms}}$ listed in Column\,9. }

\begin{figure}[H]
\centering
\includegraphics[width=0.48\textwidth, angle=0]{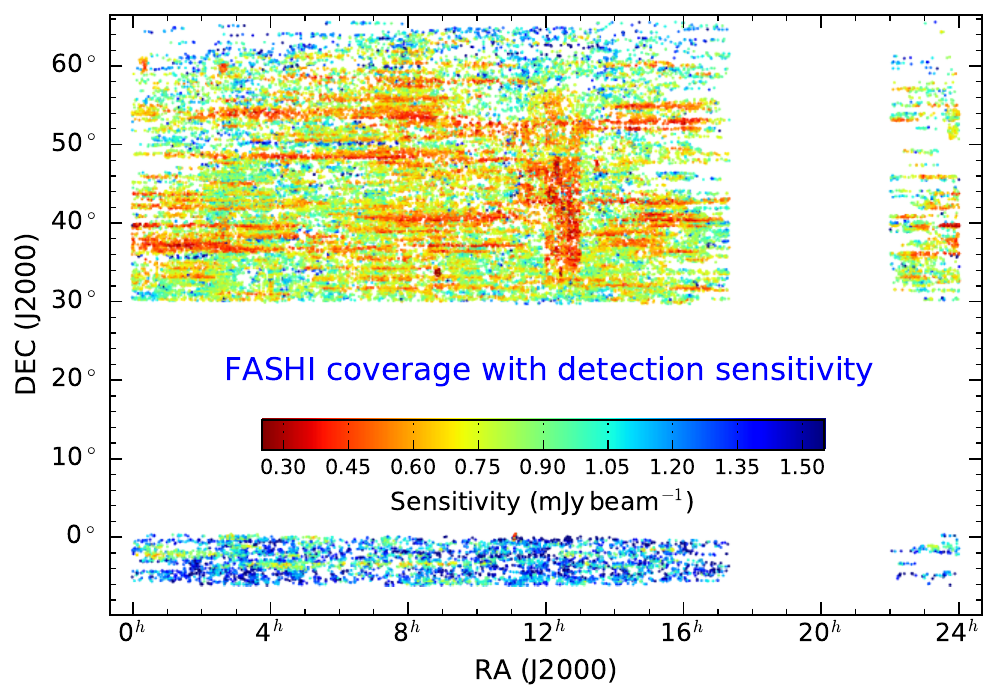}
\caption{FASHI coverage and detection sensitivity. The detection sensitivity is estimated from each detected source with the H\,{\scriptsize I} line noise ($\sigma_{\rm{rms}}$) in column\,9 and the source size ($ell_{\rm maj,\,min,\,pa}$) in column\,5 of Section\,\ref{sec:catalog_col}. It roughly shows the distribution of the map rms for each velocity channel.}
\label{Fig:sensitivity}
\end{figure}

\item {Column\,9: \HI line noise level for the spatially integrated spectral profile, $\sigma_{\rm{rms}}$ in mJy. The measurement was taken from the signal- and RFI-free sections of the integrated \HI spectrum at a spectral resolution of 6.4\,\kms. Figure\,\ref{Fig:hist_fast} displays the $\sigma_{\rm{rms}}$ histogram of FASHI sources. Furthermore, we can estimate the detection sensitivity based on the \HI line noise ($\sigma_{\rm{rms}}$) of each detected source. At a spectral resolution of 6.4\,\kms, the median detection sensitivity is $\sim$0.76\,$\mjyb$. The detection sensitivity at each FASHI source is listed in the online table, while the relevant image is shown in Figure\,\ref{Fig:sensitivity}. As shown in Figure\,\ref{Fig:sensitivity}, the detection sensitivities at different regions are currently uneven due to the constraint of the adopted schedule-filler mode. }

\item {Column\,10: Integrated \HI line flux density, $S_{\rm bf}$ in $\mjybkms$ by busy-function fitting. The SNR and log$M$, respectively in Column\,12 and 14, are calculated based on $S_{\rm bf}$ in this column rather than $S_{\rm bf}$ in Column\,11. The $S_{\rm bf}$ uncertainty is estimated with $\sigma(S_{\rm bf}) = \sqrt{N}\delta v\sigma_{\rm rms}$, where $N$ is the channel number of each spectrum only including the \HI signal, and $\delta v=6.4$\,\kms.}

\begin{figure}[H]
\centering
\includegraphics[width=0.48\textwidth, angle=0]{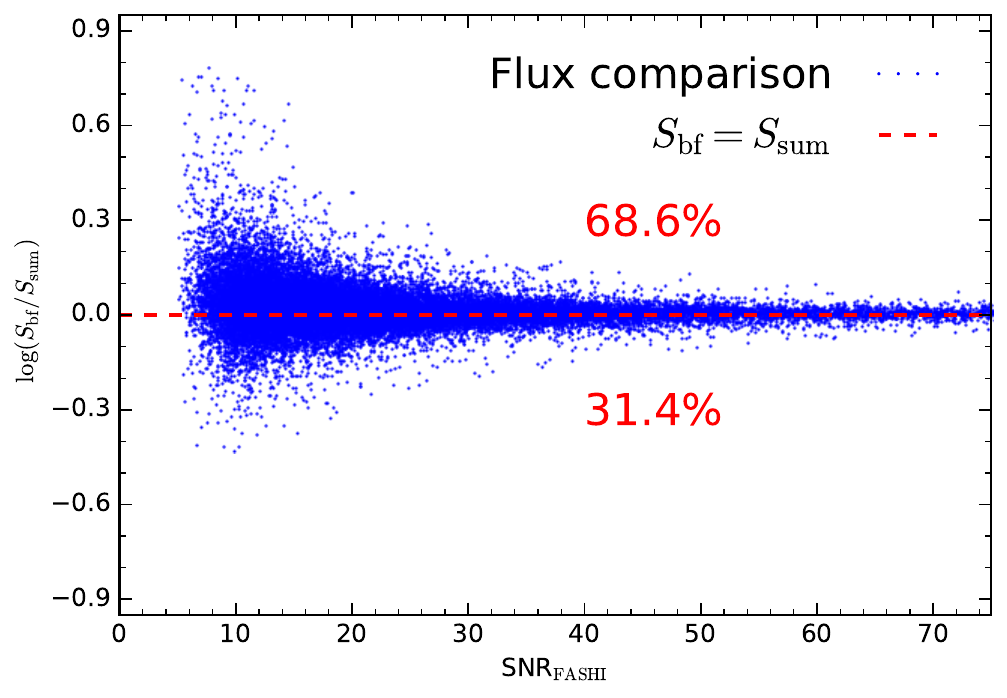}
\caption{Comparisons of the measured integrated H{\,\scriptsize I} line flux density by busy function fitting ($S_{\rm bf}$) and direct summation of all velocity channels containing signal ($S_{\rm sum}$) from each integrated spectrum. The velocity ranges are only listed in the online table. The red dashed line indicates the position of $S_{\rm bf}=S_{\rm sum}$. The percentages of occupied sources are shown below and above the red dashed line.}
\label{Fig:flux_sum_bf}
\end{figure}

\item {Column\,11: Integrated \HI line flux density, $S_{\rm sum}$ in $\mjybkms$ by summing all velocity channels containing signal within an appropriate velocity range from each integrated spectrum. Velocity ranges are only listed in the online table. Both $S_{\rm sum}$ and $S_{\rm bf}$ in Column 10 have the same level of uncertainty. Figure \ref{Fig:flux_sum_bf} illustrates the comparison of $S_{\rm sum}$ and $S_{\rm bf}$ at different signal-to-noise ratios (SNRs), revealing a noticeable systematic bias, particularly for sources with low SNR values (SNR $\lesssim$ 15). Among all sources, approximately one third exhibit measured flux densities ($S_{\rm bf}$) higher than the flux densities ($S_{\rm sum}$) obtained by summing velocity channels through busy-function fitting. The disparity among sources may have resulted from the combined effect of signal-to-noise ratio (SNR), baseline, integrated velocity range, and spectral profile, based on the examination of a large sample.  
}

\item {Column\,12: Signal-to-noise ratio (SNR) of the detection, estimated as 
   \begin{equation}
	{\rm SNR}=\left (~\frac{S_{\rm bf}}{W_{50}} \right ) \frac{w_{\rm{smo}}^{1/2}}{\sigma_{\rm{rms}}}
	\label{eq:eqsn}
	\end{equation}
where $S_{\rm bf}$ is the integrated flux density in $\mjybkms$. $w_{\rm{smo}}=W_{50}/6.4$ is a smoothing width expressed as the number of spectral resolution bins of 6.4\,\kms~bridging half of the signal width \citep[e.g.,][]{Haynes2018}. $\sigma_{\rm{rms}}$ is the rms noise across the spectrum measured in mJy at 6.4\,\kms~spectral resolution. The SNR histogram of FASHI sources is shown in Figure\,\ref{Fig:hist_fast}.}

\item {Column\,13: Distance, $D$ in Mpc. For FASHI sources with c$z_{\odot} \geq 15000$\,\kms, their distance are estimated as c$z_{\rm{cmb}}/\rm{H_0}$. Here, c$z_{\rm{cmb}}$ is the recessional velocity measured in the Cosmic Microwave Background reference frame  \citep{Lineweaver1996} and H$_0$ is the Hubble constant, which is adopted to be 75\,\kms\,Mpc$^{-1}$. For objects with $2400 \leq {\rm c}z_{\rm{\odot}} < 15000$\,\kms, we use the CF3 calculator \citep{Graziani2019,Kourkchi2020}. For objects with ${\rm c}z_{\rm{\odot}} < 2400$\,\kms, we utilized the NAM calculator \citep{Graziani2019,Kourkchi2020}. The Cosmicflows-3 Distance-Velocity Calculator employs a hierarchical Bayesian model to deduce the peculiar velocity field and matter distribution within $z\sim0.054$ using the Cosmicflows-3 galaxy distance catalog \citep{Graziani2019}. Additionally, there are 869 nearby galaxies with individual distance measurements within 11\,Mpc or corrected radial velocities $V_{\rm LG} < 600$\,\kms~\citep{Karachentsev2013}, which was applied to the distance estimate with the highest priority. The uncertainty in distance is estimated at 5\% of the distance, with a focus on the Hubble constant's uncertainty. Figure\,\ref{Fig:mass_distance} shows the distance distribution of FASHI sources.}

\item {Column\,14: Logarithm of the \HI mass, log$M$ in unit of solar mass, computed via the formula \citep[e.g.,][]{Meyer2017,Catinella2010}
   \begin{equation}
	M_{\rm HI}=\frac{2.356\times 10^5}{1+z_{\odot}} D^2 S_{\rm bf},
	\label{eq:himass}
	\end{equation}
where the adopted distance $D$ is given in Column\,13 of Table\,\ref{tab:exgalcat}, and the factor $1+z_{\odot}$ is applied for cosmological corrections \citep[e.g.,][]{Haynes2018}. The error on the mass of \HI is estimated by combining the uncertainties in distance and integrated flux density. Figure\,\ref{Fig:hist_fast} shows the mass histogram of FASHI sources.} 

\end{itemize}

\subsection{Comparison with ALFALFA sources}
\label{sec:match_alfa}

The spatial distribution of the FASHI, ALFALFA, and HIPASS sources is shown in Figure\,\ref{Fig:observed_sky}. The current FASHI project focuses on the unobserved sky that the ALFALFA survey did not cover (see also Figure\,\ref{Fig:polar}). Currently, the FASHI survey has detected a total of 41741 sources in $\sim$7600\,deg$^2$ (or $\sim$5.5 sources per deg$^2$), while the ALFALFA survey observed 31502 objects in 7000\,deg$^2$ (or $\sim$4.5 sources per deg$^2$). Although HIPASS has covered a larger area of the sky compared to FASHI, it has only detected 5374 \HI sources. The FASHI and ALFALFA mass-distance diagrams are presented in Figure\,\ref{Fig:mass_distance}. The FASHI has detected sources located at greater distances (see also Figure\,\ref{Fig:polar}) and sources with lower masses than the ones detected by ALFALFA at different distances (see also Figure\,\ref{Fig:hist_fast}). This indicates that FASHI has better detection sensitivity compared to the ALFALFA \HI survey.

Figure\,\ref{Fig:polar} shows the distribution of large-scale structures from FASHI and ALFALFA sources in different declination ranges. It covers $\rm -6^\circ\lesssim DEC \lesssim66^\circ$ and extends to a c$z_{\odot}$ range of $\sim$26323\,\kms. The FASHI data shows clearer structures from void to filament compared to the ALFALFA data. The FASHI filamentary structure is much more elongated than that of the ALFALFA in different declination spaces due to the better RA coverage in FASHI. The new FASHI data would contribute important clues to investigate the large-scale process of local universe evolution.

The histograms of the measured spectral noise level ($\sigma_{\rm rms}$), signal to noise ratio (SNR), line width ($W_{50}$ and $W_{20}$), heliocentric velocity (c$z_{\odot}$) and source mass distributions for the released FASHI and ALFALFA catalogs are displayed in Figure \ref{Fig:hist_fast}. Some useful statistical results of the FASHI sources are given in each panel. For comparison, the source mass measured by ALFALFA has been roughly corrected with the same Hubble constant as that used by FASHI. The velocity and mass histograms suggest a little lower source count for FASHI compared to ALFALFA in certain velocity and mass ($\lesssim10^8\msun$) ranges. This discrepancy may be due to the removal of 2000 candidates from the current published catalogue due to severe RFI on some frequencies and unreliable sources.  \citet{Li2022} shows that ALFALFA could underestimate the number of low HI mass galaxies below $10^9\,\msun$ due to cosmic variance, while the FASHI has detected more sources in the mass range of $10^8-10^{9}\msun$ than the ALFALFA. FASHI can detect more low mass galaxies and improve this. The FASHI will eventually provide a larger \HI population of low-mass galaxies than the ALFALFA when the 2000 candidates are confirmed with high detection sensitivities.

\begin{figure}[H]
\centering
\includegraphics[width=0.48\textwidth, angle=0]{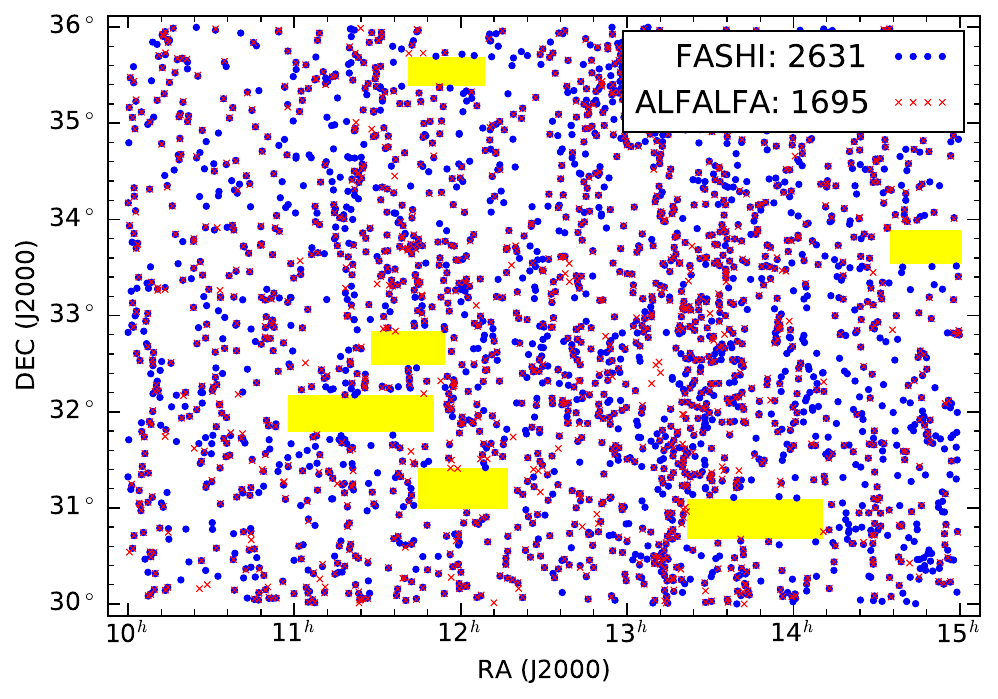}
\caption{Comparisons of all the FASHI and ALFALFA detected source distributions within the ranges of $10^h<{\rm RA}<15^h$ and $30^\circ<{\rm DEC}<36^\circ$. The yellow rectangles indicate the FASHI uncovered areas. The FASHI and ALFALFA sources are indicated with blue dot and red cross, and the source number is 2631 and 1695, respectively. }
\label{Fig:fashi_150ra225_30dec36}
\end{figure}

The comparison of all the FASHI and ALFALFA detected source distributions within the ranges of $10^h<{\rm RA}<15^h$ and $30^\circ<{\rm DEC}<36^\circ$ is shown in Figure\,\ref{Fig:fashi_150ra225_30dec36}. The FASHI and ALFALFA detected source numbers are 2631 and 1695, respectively. This indicates that within the same region, the FASHI detected source number is approximately 1.5 times higher than the ALFALFA source number. However, we must acknowledge that the current FASHI catalog also fails to detect some sources. For instance, within the ranges of $10^h<{\rm RA}<15^h$ and $30^\circ<{\rm DEC}<36^\circ$, the FASHI catalog fails to detect approximately 120 sources. This accounts for around 7.0\% of the total sources and is mainly due to three reasons. Firstly, FASHI was unable to detect faint sources due to a high level of noise in certain areas. Such unreliable and low-quality sources were removed from the current FASHI catalog. These sources require additional verification through future optical and \HI observations. Secondly, FASHI observations have not yet effectively covered some of ALFALFA's regions due to serious RFI and limited observation time. Thirdly, there is a very small number ($\lesssim$1\%) of ALFALFA catalog samples that even at high sensitivity, FASHI was unable to detect.

\begin{table*}
%\begin{sidewaystable*}
%\begin{minipage}[t]{\columnwidth}
\caption{\textbf{FASHI and ALFALFA cross-matching catalog.}}
\label{tab:cross_alfa}
\vskip 5pt
\centering \small  %\footnotesize %\scriptsize
\setlength{\tabcolsep}{1.6mm}{
\begin{tabular}{cccc|cccc}
\hline \hline
\multicolumn{4}{c|}{FASHI}  &  \multicolumn{4}{c}{ALFALFA}   \\ 
\hline
FASHI ID & RA & DEC & c$z_{\odot}$ &  AGCNr & RA & DEC & c$z_{\odot}$   \\
   &  deg & deg   &  \kms   &   & deg & deg & \kms      \\
\hline
20230013594&162.753&36.195&7131.45&5951&162.766&36.179&7109\\ 
20230013570&213.934&36.177&7408.96&245480&213.943&36.172&7427\\ 
20230013560&8.515&36.171&4809.24&104543&8.502&36.165&4809\\ 
20230013506&167.184&36.169&8246.72&216420&167.186&36.170&8177\\ 
20230058952&41.349&36.163&14472.05&124857&41.349&36.163&14466\\ 
20230054697&45.278&36.162&8158.04&132238&45.288&36.162&8207\\ 
20230052360&210.480&36.162&5123.18&245319&210.491&36.171&5141\\ 
20230013536&36.907&36.157&10740.81&1919&36.904&36.156&10755\\ 
20230013521&146.329&36.151&6555.41&190478&146.345&36.157&6537\\ 
20230013518&214.888&36.149&7419.97&9173&214.900&36.151&7426\\     
... ... \\
\hline
\end{tabular}}
\begin{flushleft}
%\begin{tabular}{l}
%\normalsize
\textbf{Notes.} The full catalog including 3620 sources could be downloaded in the online Supplementary material or at \url{https://fast.bao.ac.cn/cms/article/271/} and \url{https://zcp521.github.io/fashi}. \\
%\end{tabular}
\end{flushleft}
%\end{minipage}
%\end{sidewaystable*}
\end{table*}

\begin{figure*}
\centering
\includegraphics[width=0.45\textwidth, angle=0]{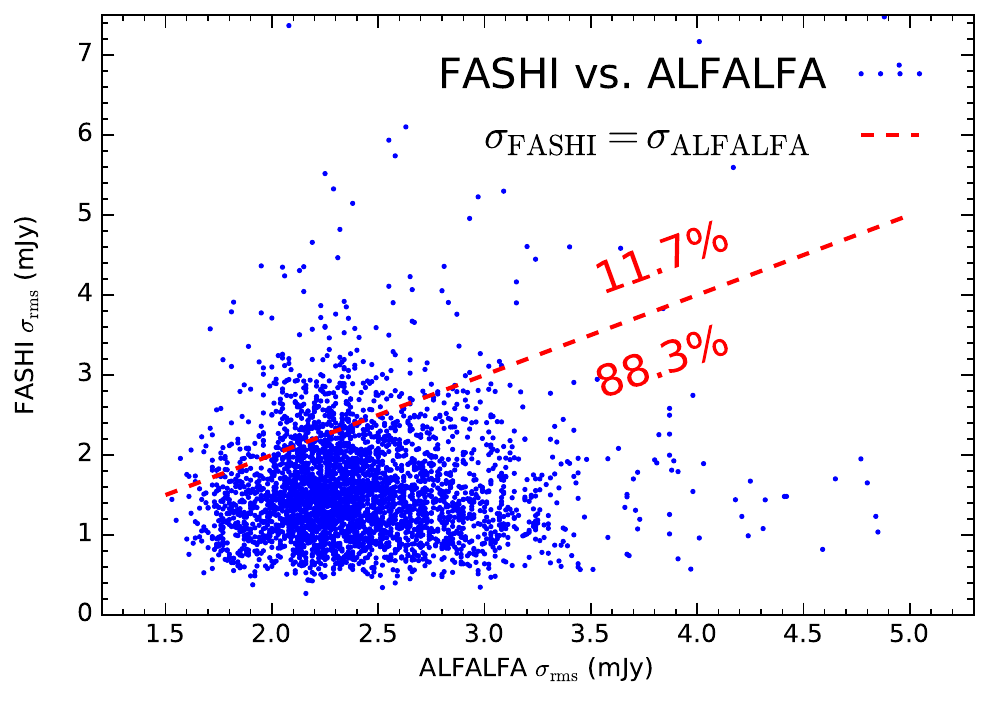}
\includegraphics[width=0.45\textwidth, angle=0]{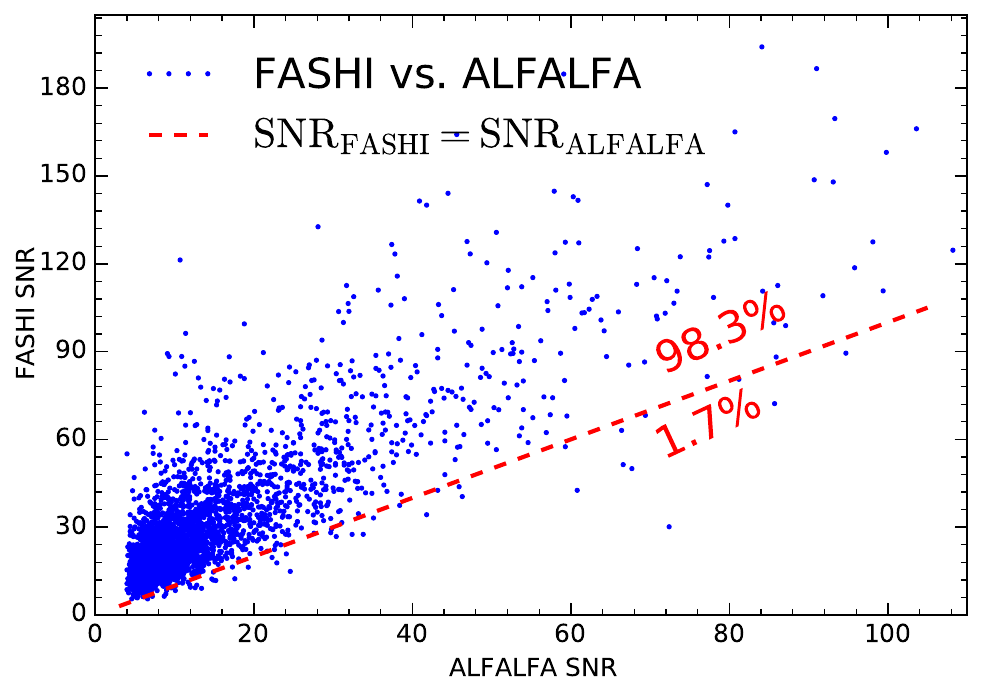}
\includegraphics[width=0.45\textwidth, angle=0]{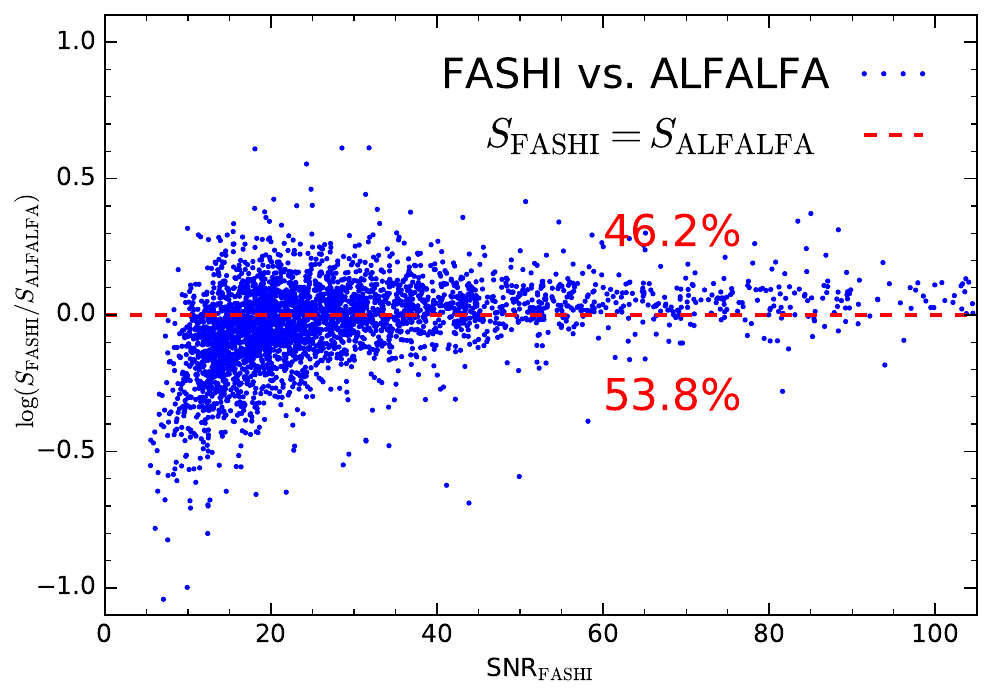}
\includegraphics[width=0.45\textwidth, angle=0]{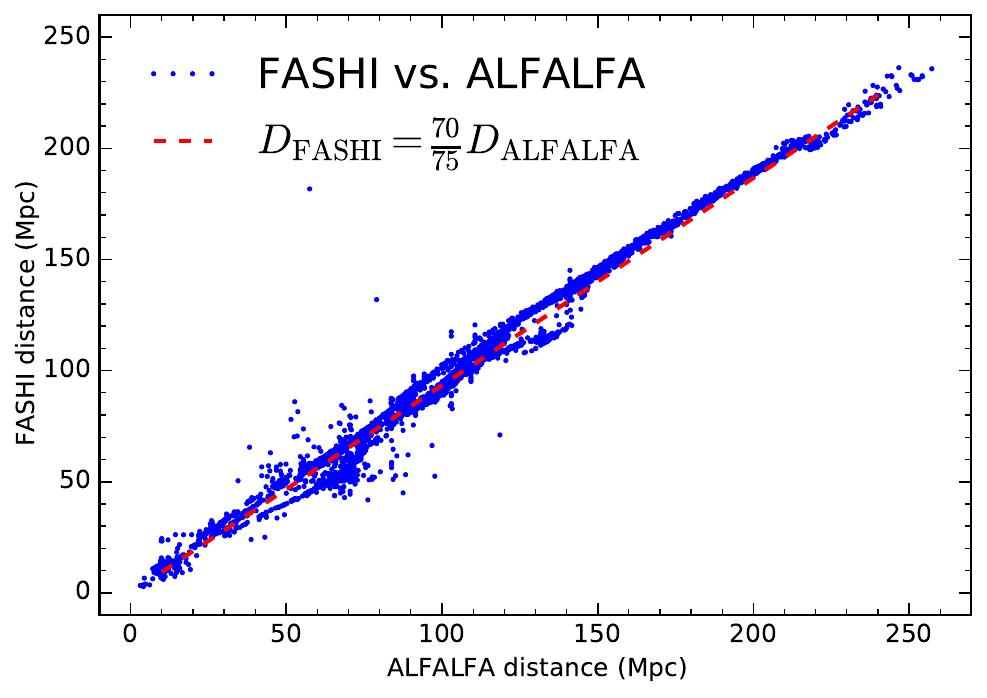}
\caption{Comparisons of the measured rms, SNR, integrated flux, and distance parameters using the cross-matched 3620 FASHI and ALFALFA sources under a condition of $\rm\delta_{RA} \leq 3'$, $\rm\delta_{DEC} \leq 3'$ and $\rm\delta_{velocity} \leq 100$\,\kms~listed in Table\,\ref{tab:cross_alfa}. The red dashed line indicates the position of the linear relation of $y_{\rm FASHI}=x_{\rm ALFALFA}$ in each panel. The percentage of source occupied is also shown in rms and SNR panels, respectively. In $\sigma_{\rm rms}$ panel, the FASHI rms is normalised to the same velocity bin with a spectral resolution of 10\,\kms, as the ALFALFA's.}
\label{Fig:fast_alfa}
\end{figure*}

Through cross-matching FASHI and ALFALFA sources while satisfying the condition of $\rm\delta_{RA} \leq 3'$, $\rm\delta_{DEC} \leq 3'$ and $\rm\delta_{velocity} \leq 100$\,\kms, we obtained a total of 3620 sources. The cross-matching results are presented in Table\,\ref{tab:cross_alfa}. The entire catalog is available for download online. This sample is sufficiently large for comparing the parameters between FASHI and ALFALFA sources. Figure\,\ref{Fig:fast_alfa} displays the comparison of the measured rms, SNR, integrated flux, and distance parameters between FASHI and ALFALFA sources using the cross-matched sample. According to Figure\,\ref{Fig:fast_alfa}, it can be observed that a lower noise level is found in 88.3\% of FASHI sources, and 98.3\% of FASHI sources have higher SNR than cross-matched ALFALFA sources with the same spectral resolution of 10\,\kms. Furthermore, we find that the integrated flux distributions of the two survey sources are in good agreement. We also compared the integrated \HI lines of these bright sources ($\gtrsim7.0\,\jybkms$) with those obtained from ALFALFA. The FASHI's integrated bright point sources show a flux excess of approximately 10\% compared to the ALFALFA sources for sources with $\rm SNR_{FASHI}\gtrsim40$. Perhaps the FASHI sources, due to their extremely high sensitivity, include more tidal clouds and can provide a more accurate measurement for bright sources than the ALFALFA according to \citet{Hoffman2019AJ}. For sources with relatively low SNR ($\rm SNR_{FASHI}\lesssim20$), the FASHI's measured flux distribution is lower than that of the ALFALFA's. Sources with low SNR are typically associated with very faint sources. Due to 98.3\% of FASHI sources having better SNR compared to ALFALFA sources, the latter will have significantly lower SNR than the faint sources in FASHI. Measuring a very faint source often leads to high levels of uncertainties. It is possible that FASHI provides more accurate measurements than ALFALFA for faint sources \citep[see also analysis in][]{Jing2023}.

In addition, the measured distances in Figure\,\ref{Fig:fast_alfa} show a consistent relationship after correction with $D_{\rm FASHI} = \frac{70}{75}D_{\rm ALFALFA}$ between FASHI and ALFALFA, because ALFALFA source distances are estimated using a different Hubble constant H$_{0}$ = 70\,\kms Mpc$^{-1}$ from FASHI (see details for distance calculation in Sections\,\ref{sec:distances} and \ref{sec:catalog_col}). This further suggests that the released FASHI catalogs are highly reliable.

\subsection{Optical counterparts}
\label{sec:oc}

Due to the discrepancy between the spatial resolution of the optical image (typically 1$''$) and the large beam size ($\sim$2$\dotmin$9) of FAST, it may be difficult to detect trustworthy optical counterparts, especially for merging systems where only one \HI target is present in the FASHI catalog but there are several optical galaxies at similar redshifts. Furthermore, the long extended \HI tail of galaxies can be regarded as one \HI bright target, which lacks clear optical counterparts. In order to identify reliable optical counterparts, conducting high-resolution \HI follow-up observations is necessary. Nonetheless, this is excessively costly for large blind \HI survey projects like FASHI. To gain an initial understanding of the FASHI galaxy properties, we cross-matched the FASHI targets with the findings of optical galaxy surveys.

To identify bright galaxy counterparts that may be saturated or masked in wide-field galaxy surveys such as the SDSS, we first compared the FASHI catalog with the local galaxy sample Siena Galaxy Atlas\footnote{\url{https://www.legacysurvey.org/sga/sga2020/}} (SGA), which contains 383620 galaxies with $D_{25} > 20''$. For each FASHI \HI target, we searched for SGA counterparts under the condition $\rm\delta_{RA} \leq 3'$, $\rm\delta_{DEC} \leq 3'$ and $\rm\delta_{speed} \leq 1000$\,\kms. If more than one candidate galaxy was present, we selected the optical counterpart with the shortest distance. From the SGA we obtained 14072 counterparts, which are listed in Table\,\ref{tab:cross_sga}. For the other 27669 \HI targets, we cross-matched with the SDSS spectroscopic redshift catalog \citep{Abazajian2009} using the same strategy as for the SGA cross-match, yielding 2900 targets, which are listed in Table\,\ref{tab:cross_sdss}. In total, we found 16972 optical counterparts from the SGA and SDSS spectroscopic redshift catalogs.

Figure\,\ref{Fig:hist_match_sdss} reports the measured position and velocity separations along with statistical results of the cross-matched sample of 16972 sources from the FASHI and SGA \& SDSS projects. The statistical results indicate that the position and absolute velocity separations of 87.7\% and 87.6\%, respectively, of the FASHI sources are less than 60$''$ and 50\,\kms~compared to the cross-matched SGA \& SDSS sources. The small offset between the FASHI sources and the optical counterparts suggests that the nearest candidates are statistically appropriate for our purpose. Both separations are consistent with the cross-match results obtained by the SGA and SDSS. The SGA and SDSS spectroscopic data are deemed highly reliable. This observation implies that FASHI sources have more accurate position and velocity measurements.

To obtain complete optical counterparts for all FASHI targets, we performed a further cross-match with the SDSS photometric catalog using CasJobs\footnote{\url{http://mastweb.stsci.edu/mcasjobs/}} to identify all SDSS galaxies within 3$'$ of each FASHI target. For each SDSS galaxy within the FAST beam (assuming the r-band Petrosian magnitude $m_r$ and the distance to the \HI center is $D_i$ arcsec), we considered the probability of the serendipitous counterparts based on the surface number density in the r band, which is $P_{\rm mag} = e^{-\pi D_i^2 n (m<m_r) }$, where $n (m<m_r)$ is the surface number density brighter than $m_r$ \citep{Downes1986,Zhangl2022}. If the SDSS target is faint, the expected number of galaxies that randomly appear in the area of $\pi D_i^2$ would be high. Therefore, such a faint galaxy would be more likely to be randomly associated with the \HI target and, thus, less likely to be the counterpart. Moreover, the distribution of the RA and DEC offset also shows that it is more likely to find counterparts with less center offset. Therefore, we treated the offset as a probability $P_{\delta \rm RA} \times P_{\delta \rm DEC}$, assuming a Gaussian distribution of the offsets, with the center and $\sigma$ from the SGA offset results. Thus, the optical counterpart probability ($P_{\rm OC}$) of each SDSS galaxy within the beam is $P_{\rm OC} = P_{\rm mag} \times P_{\delta \rm RA} \times P_{\delta \rm DEC}$. We emphasize that this is a relative probability used to filter out unlikely optical counterparts within the FAST beam.

Using the probability method above, we will always find an SDSS target as the most likely counterpart to the \HI target. However, for the optically faint galaxies \citep{2017ApJ...842..133L, Xujl2023, 2023AJ....165..197G}, the SDSS image is not deep enough to detect the targets, or the low surface brightness galaxies are not detected by the SDSS pipeline. On the other hand, for a \HI target with SDSS coverage but no SDSS spec-$z$ counterparts, the optical counterpart would be fainter than $m_{r} \sim 17.7$. Finally, cross-matching with SDSS photometric data yielded 10975 optical counterparts. The derived results of the cross-match with the photometric catalog are listed in Table\,\ref{tab:cross_sdss_phot}.

The FASHI coverage includes part of the Galactic plane where there is little SDSS coverage. Although these regions ($20^\circ<\rm RA<120^\circ$) are covered by the Pan-STARRS project, the crowded foreground star overlap and the strong dust extinction from the Milky Way limit the accuracy of the Galactic photometry. In this paper we have only matched the FASHI targets with SGA and SDSS, and have not identified optical counterparts for the remaining 13794 FASHI targets.

\begin{table*}
\caption{\textbf{The identified optical counterparts from the SGA catalog.}}
\label{tab:cross_sga}
\vskip 5pt
\centering \small  %\footnotesize %\scriptsize
\setlength{\tabcolsep}{1.6mm}{
\begin{tabular}{cccc|cccc}
\hline \hline
\multicolumn{4}{c|}{FASHI}  &  \multicolumn{4}{c}{SGA}   \\ 
\hline
FASHI ID  &  RA & DEC & $z_{\odot}$ & SGA & RA & DEC & $z_{\odot}$ \\
&     deg & deg &  & &   deg & deg &       \\
\hline
20230000823&166.736&-6.193&0.02269&PGC1035830&166.738&-6.183&0.02290\\ 
20230042079&165.515&-6.178&0.03145&PGC1035861&165.504&-6.180&0.03160\\ 
20230060002&356.034&-6.162&0.00704&PGC072252&356.044&-6.170&0.00704\\ 
20230000831&58.707&-6.171&0.03664&PGC1036048&58.703&-6.163&0.03644\\ 
20230000832&169.267&-6.141&0.04476&PGC1036233&169.281&-6.148&0.04532\\ 
20230000833&171.782&-6.140&0.02999&PGC035230&171.789&-6.139&0.02995\\ 
20230000834&168.912&-6.137&0.01945&PGC1036364&168.917&-6.137&0.02020\\ 
%20230000840&51.378&-6.135&0.03468&PGC012803&51.381&-6.129&0.03481\\ 
... ... \\
\hline
\end{tabular}}
\begin{flushleft}
%\begin{tabular}{l}
%\normalsize
\textbf{Notes.} The complete catalog including 14072 sources could be downloaded in the online Supplementary material or at \url{https://fast.bao.ac.cn/cms/article/271/} and \url{https://zcp521.github.io/fashi}. \\
%\end{tabular}
\end{flushleft}
%\end{minipage}
%\end{sidewaystable*}
\end{table*}

\begin{table*}
%\begin{sidewaystable*}
%\begin{minipage}[t]{\columnwidth}
\caption{\textbf{The identified optical counterparts from the SDSS spectroscopic catalog.}}
\label{tab:cross_sdss}
\vskip 5pt
\centering \small  %\footnotesize %\scriptsize
\setlength{\tabcolsep}{1.6mm}{
\begin{tabular}{cccc|cccc}
\hline \hline
\multicolumn{4}{c|}{FASHI}  &  \multicolumn{4}{c}{SDSS}   \\ 
\hline
FASHI ID & RA & DEC & $z_{\odot}$ &  OBJID & RA & DEC & $z_{\odot}$   \\
   &  deg & deg   &     &   & deg & deg &       \\
\hline
20230053284&48.425&-6.073&0.03144&1237649964065161287&48.429&-6.084&0.03125\\ 
20230060070&37.368&-5.847&0.04766&1237679439889301650&37.364&-5.848&0.04810\\ 
20230060087&36.523&-5.777&0.03241&1237679439888908518&36.530&-5.781&0.03209\\ 
20230060106&35.641&-5.708&0.03285&1237679341105971489&35.638&-5.713&0.03276\\ 
20230001011&54.560&-5.652&0.02203&1237652901845598378&54.566&-5.659&0.02183\\ 
20230001046&54.542&-5.535&0.02211&1237652901845598221&54.554&-5.540&0.02215\\ 
20230060210&58.242&-5.077&0.01724&1237652901847171254&58.241&-5.080&0.01735\\ 
%20230053259&35.945&-4.857&0.04396&1237679253598372170&35.959&-4.843&0.04378\\ 
... ... \\
\hline
\end{tabular}}
\begin{flushleft}
%\begin{tabular}{l}
%\normalsize
\textbf{Notes.} The complete catalog including 2900 sources could be downloaded in the online Supplementary material or at \url{https://fast.bao.ac.cn/cms/article/271/} and \url{https://zcp521.github.io/fashi}. \\
%\end{tabular}
\end{flushleft}
%\end{minipage}
%\end{sidewaystable*}
\end{table*}

\begin{table*}
%\begin{sidewaystable*}
%\begin{minipage}[t]{\columnwidth}
\caption{\textbf{The identified optical counterparts from the SDSS photometric catalog.}}
\label{tab:cross_sdss_phot}
\vskip 5pt
\centering \small  %\footnotesize %\scriptsize
\setlength{\tabcolsep}{1.6mm}{
\begin{tabular}{cccc|cccc}
\hline \hline
\multicolumn{4}{c|}{FASHI}  &  \multicolumn{4}{c}{SDSS}   \\ 
\hline
FASHI ID & RA & DEC & $z_{\odot}$ &  OBJID & RA & DEC & $P_{\rm mag}$   \\
   &  deg & deg   &     &   & deg & deg &       \\
\hline
20230042080&125.705&-6.169&0.04885&1237674460407923463&125.731&-6.163&0.848\\ 
20230042081&126.961&-6.160&0.02162&1237673708252234978&126.971&-6.155&0.936\\ 
20230000830&172.136&-6.156&0.01670&1237671142014648688&172.146&-6.146&0.994\\ 
20230000845&44.275&-6.107&0.03168&1237679438818574478&44.286&-6.143&0.969\\ 
20230000841&122.921&-6.115&0.03842&1237673710935081236&122.928&-6.125&0.996\\ 
20230060006&191.710&-6.096&0.01264&1237671763173310839&191.721&-6.116&0.860\\ 
20230042084&46.376&-6.103&0.02957&1237679339500011708&46.377&-6.101&0.999\\ 
%20230060012&207.360&-6.081&0.00506&1237671955911802962&207.364&-6.088&0.984\\ 
... ... \\
\hline
\end{tabular}}
\begin{flushleft}
%\begin{tabular}{l}
%\normalsize
\textbf{Notes.} The complete catalog including 10975 sources could be downloaded in the online Supplementary material or at \url{https://fast.bao.ac.cn/cms/article/271/} and \url{https://zcp521.github.io/fashi}. \\
%\end{tabular}
\end{flushleft}
%\end{minipage}
%\end{sidewaystable*}
\end{table*}

\begin{figure}[H]
\centering
\includegraphics[width=0.48\textwidth, angle=0]{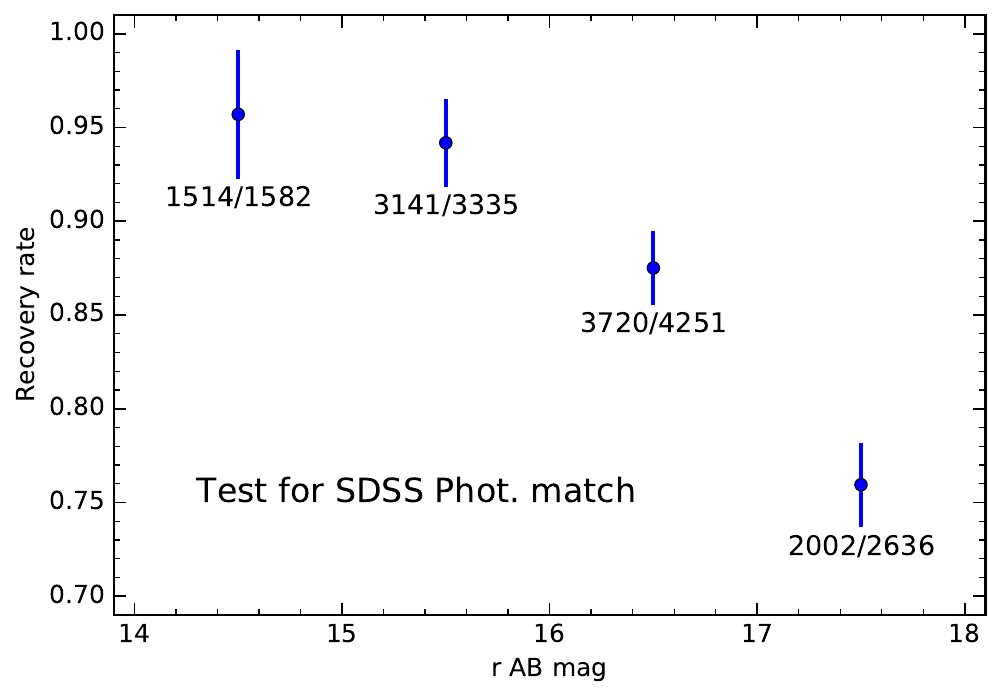}
\caption{The corresponding recovery rate from the photometry matching results (listed in Table\,\ref{tab:cross_sdss_phot}) having both the SDSS spectroscopic and photometric data, as well as the r-band cmodel magnitude from the SDSS. The recovery rate is high for bright targets, but drops to about 80\% at the faint end. Unfortunately, we do not have the recovery rate for the counterparts only from the SDSS photometric catalog, but we hope that the future spectroscopic redshift survey projects, such as DESI and PFS, will help to identify more optical counterparts.}
\label{Fig:recovery_rate}
\end{figure}

To validate our matching method, we applied it to match the FASHI \HI catalog to the SDSS spectroscopic redshift and SDSS photometric catalogs. We used the same methods as for the SGA catalog match and the SDSS photometry match. Since the spectroscopic redshift results are more reliable for the optical counterparts, we selected sources that have both photometry data and spectroscopic redshift. We then compared the photometry matching results with the spectroscopic redshift catalog. The derived distribution of the recovery rate of the counterparts is shown in Figure\,\ref{Fig:recovery_rate}. We found that the recovery rate is about 80\% for targets with $r\simeq 17$ AB mag, and the rate decreases for fainter targets.

\begin{figure}[H]
\centering
\includegraphics[width=0.48\textwidth, angle=0]{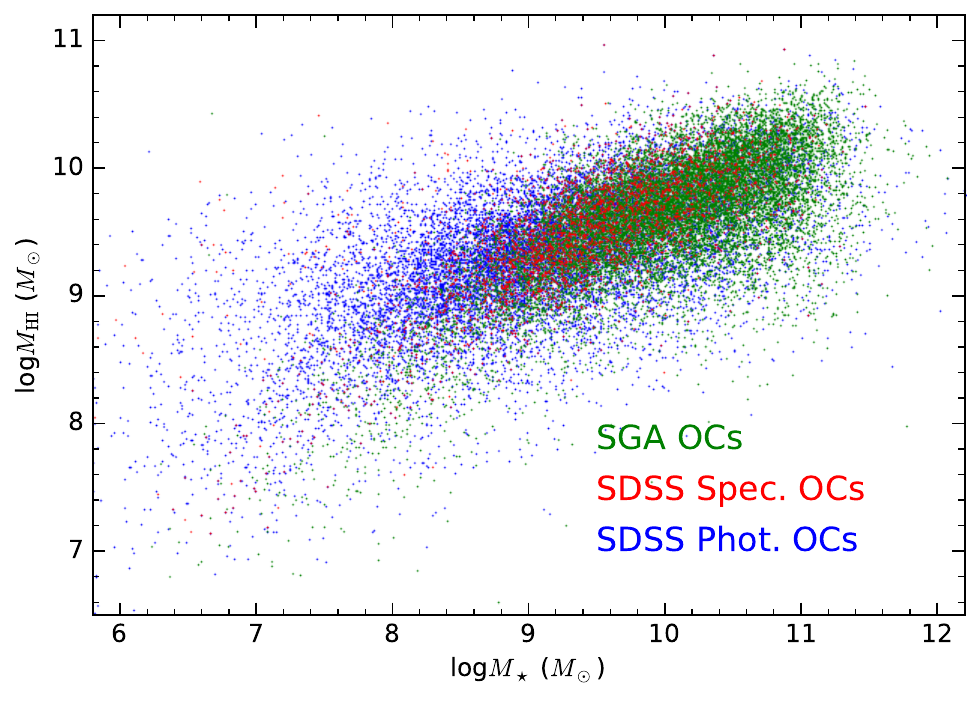}
\caption{The H\,{\scriptsize I} mass and the stellar mass of the counterparts distribution. The SGA catalog counterparts are in green, and the SDSS spectroscopic and the photometric counterparts are in red and blue, respectively.}
\label{Fig:mass_mass}
\end{figure}

To provide a first overview of the optical properties of the FASHI sample, we present the \HI and stellar mass diagram in Figure\,\ref{Fig:mass_mass}. The stellar mass was estimated using the mass-to-light ratio from \citet{2003ApJS..149..289B}, with a factor of 1.6 applied to convert the Salpeter initial mass function (IMF) to the Chabrier IMF \citep{2003PASP..115..763C}. As expected, the SGA counterparts are associated with more massive galaxies, while the SDSS photometric counterparts are mainly associated with less massive galaxies.

Since the SDSS photometric counterparts are mainly associated with low-mass galaxies, whose \HI mass is also subject to detection reliability, it may be more effective to use future spectroscopic redshift survey projects, such as DESI \citep{DESI2022} and PFS \citep{Wang2020}, to identify fainter optical counterparts and explore their optical properties, rather than relying solely on improving the algorithm for selecting optical counterparts from the photometric catalog.

\subsection{Extragalactic OH megamasers}\label{sec:ohmaser}

OH Megamasers (OHM) are powerful sources of line emission detected at L band, originating from the starburst nuclei in merging galaxy systems \citep{Darling2002,Darling2006,Giovanelli2015}. Approximately 130 such sources have been discovered to date \citep[e.g.,][]{Darling2002,Haynes2018,Roberts2021,Suess2016}. OHM lines shifted from their rest frequencies, which are 1612.231, 1665.402, 1667.359, and 1720.530\,MHz, into the targeted \HI bandpass may contaminate blind \HI line surveys of the local universe. The FAST with its Ultra-Wide Bandwidth (UWB) receiver can simultaneously cover frequencies ranging from 500 to 3300\,MHz and is expected to be an influential instrument for detecting OH emission from the Milky Way and extragalaxies \citep{Zhang2023}. For FASHI, the redshift range of the 1667.359\,MHz OHM line corresponds to 0.150 $< z <$ 0.278 within a frequency range of 1305-1450\,MHz. Essentially, the extragalactic \HI and OHM lines exhibit similar spectral profiles at various redshifts. Moreover, the 1612.231 and 1720.530\,MHz lines are consistently too weak to be recognized by FASHI, and the 1665.402 and 1667.359\,MHz lines blend together due to their broad line width. Therefore, it is impossible to distinguish them based solely on the line profile for large sample identifications. In cases where optical redshifts are provided, the OHM can be easily identified among the extracted \HI candidates (refer to the OHM sample in Figure\,\ref{Fig:fast_sample}). Alternatively, we are considering employing WISE infrared colours and magnitudes to distinguish OHMs from \HI line emitters \citep[refer to][]{Suess2016}. The identified OHM candidates with optical redshifts have been removed from the FASHI catalog, and their details will be presented in a forthcoming work. It is likely to that there still exist OHM sources remaining in the current FASHI catalog.

\subsection{Radio recombination lines}\label{sec:rrl}

\begin{figure*}
\centering
\includegraphics[width=0.49\textwidth, angle=0]{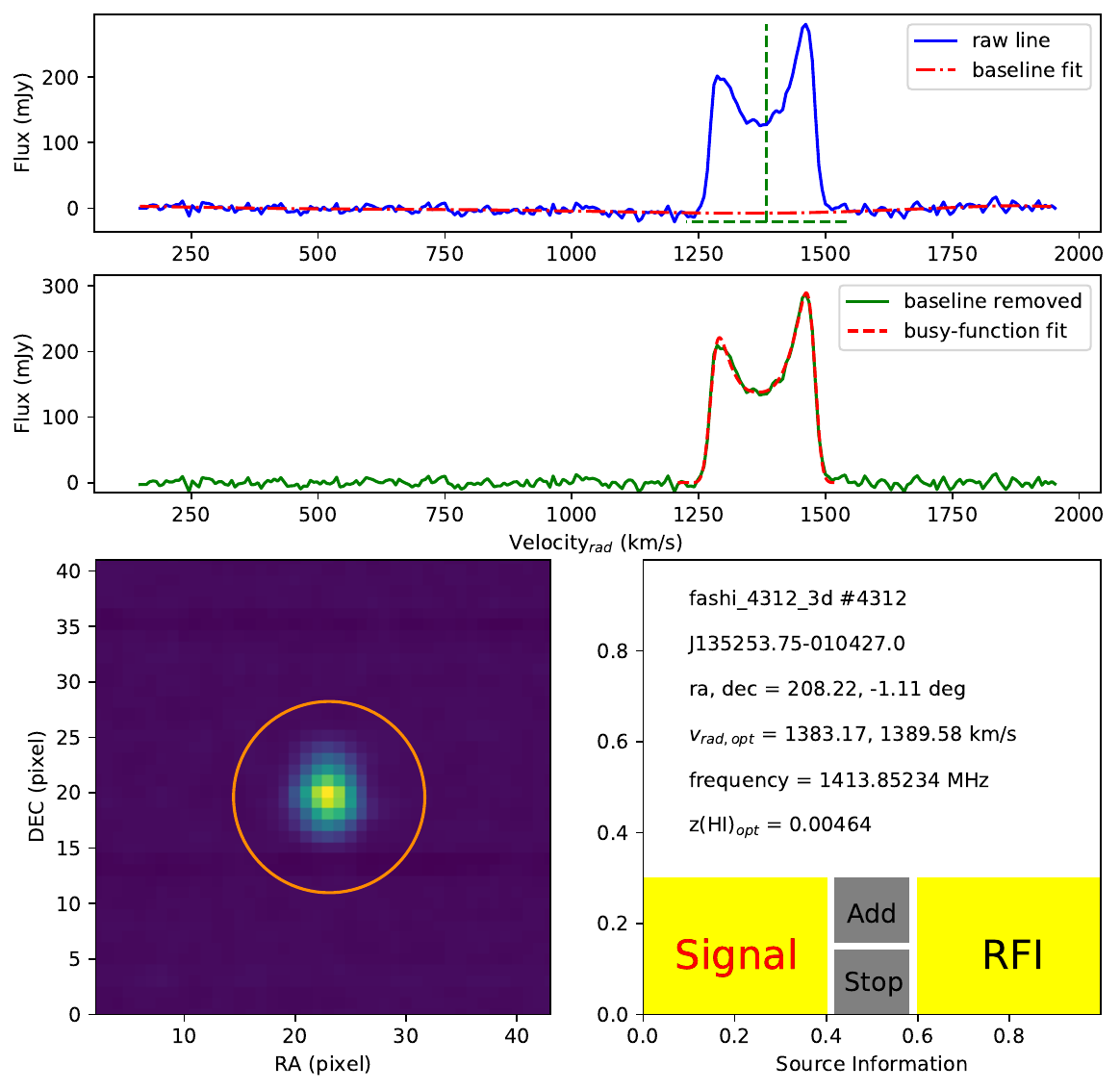}
\includegraphics[width=0.49\textwidth, angle=0]{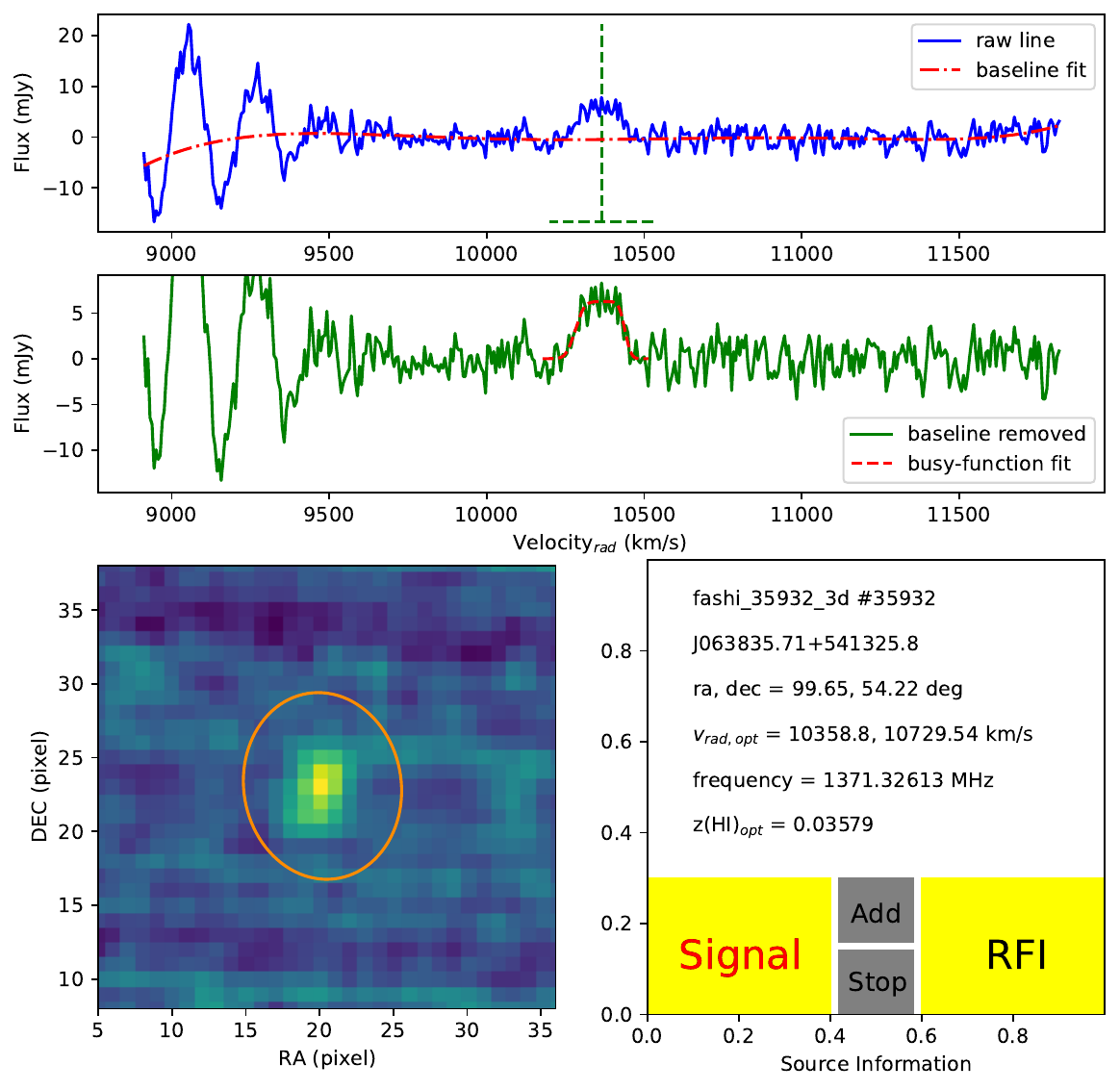}
\includegraphics[width=0.49\textwidth, angle=0]{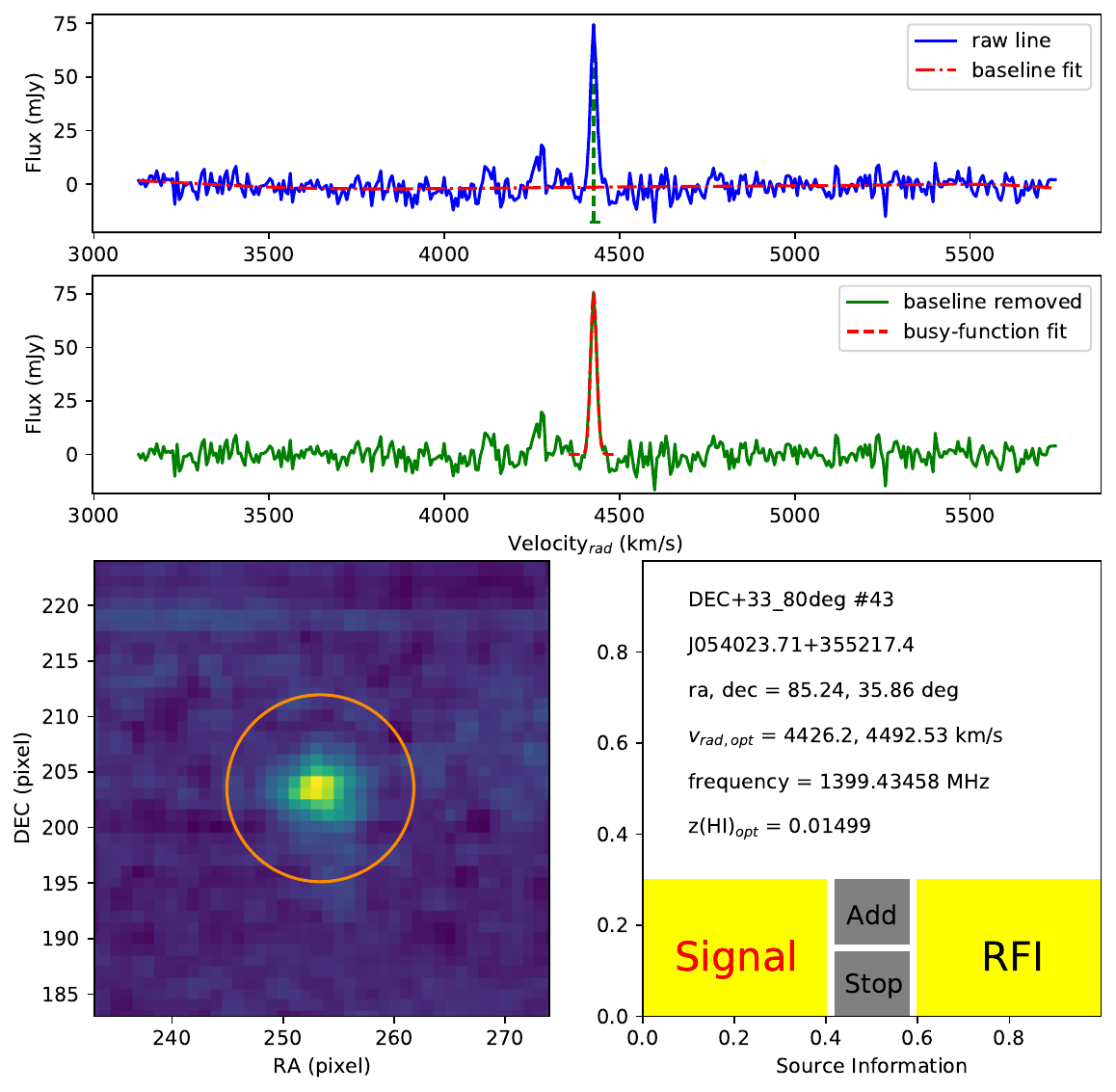}
\includegraphics[width=0.49\textwidth, angle=0]{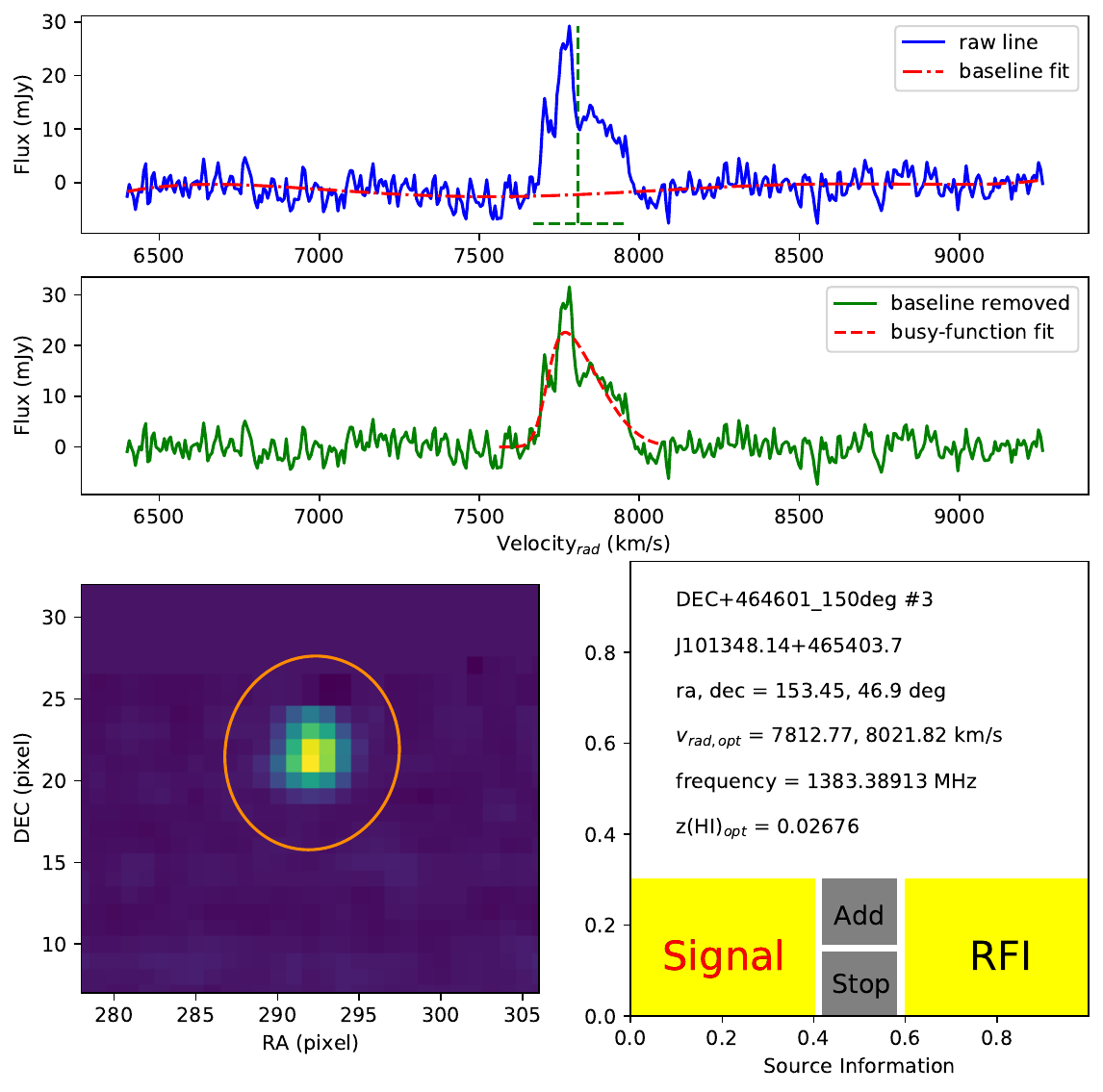}
\caption{Interactive interface for the FASHI source finder. In each panel, the upper blue spectrum represents the raw line extracted from data fits cube produced from the \texttt{HiFAST} pipeline \citep{Jing2023}, the dash-dot line displays the fitted baseline, while the vertical and horizontal dashed green lines roughly indicate the line center velocity and line width; the middle green spectrum is the baseline removed data, and the red line represents busy-function fit; the lower-left image displays the moment 0 map; the lower-right section presents the source information, assuming all the emission lines are H{\,\scriptsize I}. Top left: a normal double-horn galaxy. Top right: a normal galaxy with RFI, located at the around 9000-9500\,\kms. Bottom left: a radio recombination line. Bottom right: a new recently discovered OHM (Zhang et al. in prep.) that corresponds to IRAS\,10106+4708 with a redshift of $z_{\rm OH}=0.20586$ \citep{Abazajian2009}. }
\label{Fig:fast_sample}
\end{figure*}

Radio recombination lines (RRLs) originate from gas ionized by young massive stars in the Galactic \HII regions \citep{Zhang2014,Zhang2021,Hou2022}. Galactic RRLs typically have narrower line widths (FWHM $<$ 50\,\kms) than most extragalactic \HI spectral lines (see RRL sample in Figure\,\ref{Fig:fast_sample}). The frequency band of FAST (1050-1450\,MHz) can simultaneously cover at most 23 RRLs for H$n\alpha$, C$n\alpha$, and He$n\alpha$ (principal quantum number $n=164-186$), respectively. According to \citet{Zhang2021}, 21 RRLs of H$n\alpha$ and C$n\alpha$ at 1050-1450\,MHz have been simultaneously detected with strong emission signals. The FASHI coverage towards the Galactic direction can aid in building a more expansive sample of RRLs from the Milky Way. The RRLs from the FASHI data have been identified using the criteria that the same spatial position should have consistent flux density and line width at different transition frequencies. The line frequencies of 1326.792, 1350.414, 1399.368, and 1374.600\,MHz are relevant within the frequency range of the currently released data (1305.5-1419.5\,MHz). As the first released FASHI catalog concentrates on the extragalactic \HI source, we have excluded the Galactic RRL targets from the catalog.

\subsection{Galactic high velocity cloud}\label{sec:hvc}

Galactic high velocity clouds (HVCs) are gas clouds located nearby the Galactic plane, which move at speeds ($\gtrsim90$\,\kms) substantially different from the rotation of the disk of the Milky Way galaxy \citep{Adams2013,Adams2016,Moss2013,Verschuur1973}. HVCs are thought to be clouds of material falling into the Galaxy from the outside the Galactic plane \citep{Wakker1997}. Additionally, HVCs might potentially have extragalactic origins and they may be very similar to the low-mass population of failed galaxies. HVCs share some features with the nearby extragalactic sources \citep{Haynes2018}. And it is not possible to identify any stellar counterparts in HVCs. Therefore, the first released FASHI catalog includes the HVCs, although these are not identified.

\subsection{Accessing FASHI data}\label{sec:spectra}

In addition to the catalog listed in Table\,\ref{tab:exgalcat}, the corresponding integrated 1D \HI line spectra and 2D moment maps for all sources have been extracted from the 3D fits cube over at least a 400 velocity channel window. The entire FASHI \HI catalog will be made directly available to any user upon publication of this paper. The reduced 1D cubelets may be accessed reasonably by contacting the corresponding authors. The reduced 2D and 3D data cubes are currently provided only to internal users. Furthermore, all the raw data of the FASHI project will be freely accessible to any user after a twelve-month privileged period\footnote{The twelve-month privileged period is after each scan observation, not after the entire survey is completed.} as per the FAST data release policy\footnote{See details in \url{https://fast.bao.ac.cn/cms/article/129/}}.

\section{Discussion}\label{sec:discuss}

\subsection{Reliability}

The released catalog (Table\,\ref{tab:exgalcat}) contains sources extracted using \texttt{SoFiA} at a 4.5$\sigma$ threshold. The sources were inspected through interactive manual source extraction for reliability, based on the 0th, 1st, and 2nd moments, integrated spectral profile, and SNR ($\rm SNR > 5.0$). Human intervention was used to optimize the measurement accuracy and improve the catalog's reliability by rejecting spurious detection that corresponds to low-level RFI, poorly sampled data, and residual baseline fluctuations. The optical counterpart of each source was identified and listed in Tables\,\ref{tab:cross_sga}, \ref{tab:cross_sdss}, and \ref{tab:cross_sdss_phot} (see Section\,\ref{sec:oc} for details). Sources located near the cube's edge, blanked pixels, or frequency were removed. As a result, all published FASHI sources are of relatively high confidence.

\subsection{Completeness}

In the first released FASHI \HI catalog, the sensitivities for detection displayed in Figure \ref{Fig:sensitivity} are uneven across different regions. This is because FASHI could only use the schedule-filler time, as detailed in Section \ref{sect:strategy}. As a result, the completeness becomes relatively low at a high detection sensitivity or at a relatively high redshift. Additionally, the strange gaps of the FASHI targets at $\sim$190, $\sim$210 and $\sim$266\,Mpc are also resulted in a relatively low completeness. Moreover, approximately 2000 sources, which have low reliability due to low signal-to-noise ratio ($<5.0$) or being located at the edge of velocity (channel) or spatial (pixel), have been removed from FASHI catalog and need more time for identification. Thus, if completeness analysis is needed, the completeness levels should be estimated at different detection sensitivity levels. 

\subsection{Caveats and comments}

Despite the high quality of most FASHI data, caution should be exercised when using the source catalog presented in Table\,\ref{tab:exgalcat}. Some of caveats and comments are listed below:

\begin{itemize}

\item {Caveats: In some double peak spectra in FASHI data, the W$_{50}$ values measured by busy-function are multiple times lower than the W$_{20}$ values when the lower peak intensity is less than 50\% of the higher peak intensity. If using such W$_{50}$ values for calculations and statistics, they should be further corrected. There are approximately 2000 sources that have low reliability due to low signal-to-noise ratio ($<5.0$) or being located at the edge of velocity (channel) or spatial (pixel). Low-reliability sources have been excluded from the current FASHI catalog. Future optical and \HI observations need to be conducted to ensure their accuracy. Some negative sources (absorption lines) have been detected, but they are not included in this released catalog. More time is required for the data reduction of the negative \HI sources.}

\item {The FASHI project covers certain regions of the Milky Way galaxy. Despite our best efforts, some of the Galactic RRLs might still be present in the catalog. Additionally, the extragalactic OHMs have been excluded from the catalog only if the SDSS catalog contains a matching known spectroscopic redshift. 
}
\end{itemize}

\section{Summary}\label{sec:summ}

The FASHI project intends to scan the entire sky visible from the FAST observatory, which covers an area of approximately 22000 square degrees across a declination range of $-14\degree$ to $+66\degree$. The FASHI project is expected to detect over 10$^5$ \HI sources across the portion of the FAST sky  with a frequency range of 1050-1450\,MHz. This detection was possible up to a redshift of $z \approx 0.35$. FASHI has surveyed an area of approximately 7600 square degrees from August 2020 to June 2023. The median detection sensitivity achieved was $\sim$0.76\,$\mjyb$ at a resolution of $\sim$6.4\,\kms~at $\sim$1.4\,GHz. The frequency range used for the search was approximately 1305.5-1419.5\,MHz. In total, 41741 \HI sources were detected and identified. Our detection rate is $\sim$5.5 sources per deg$^2$, which is slightly higher than that of the ALFALFA survey. This is thanks to the higher sensitivity of FAST. FASHI is sampling a wide range of hosts galaxies from local, very low \HI mass dwarfs to gas-rich massive galaxies seen to $z\sim0.09$. The FASHI survey provides the largest extragalactic \HI catalog, offering an unbiased view of \HI content and structures in the universe. Up to now, FASHI is the most sensitive, comprehensive, and previously unheard of blind extragalactic \HI survey of the FAST sky, with a higher spectral and spatial resolution, and broader coverage than the ALFALFA. We cross-matched SGA and SDSS spectroscopic redshift data, together with SDSS photometric data, to identify optical counterparts for 27947 FASHI sources. Approximately 59.3\% of the complete FASHI catalog lacks coverage by any spectroscopic redshift survey. This presents a large number of candidates for future spectroscopic redshift and photometric surveys, such as DESI and PFS. 

%%%%%%%%%%%%%%%%%%%%%%%%%%%%%%%%%%%%%%%%%%%%%%%%%%%%%%%
%%% Acknowledgements. ??
%%%%%%%%%%%%%%%%%%%%%%%%%%%%%%%%%%%%%%%%%%%%%%%%%%%%%%%
\section*{Acknowledgements}

This work is supported by the National Key R\&D Program of China (Nos.\,2018YFE0202900 and 2022YFA1602901). CPZ acknowledges support by the West Light Foundation of the Chinese Academy of Sciences (CAS). CC is supported by the National Natural Science Foundation of China, Nos.\,11803044, 11933003, and 12173045. This work is sponsored (in part) by the Chinese Academy of Sciences (CAS), through a grant to the CAS South America Center for Astronomy (CASSACA). We acknowledge the science research grants from the China Manned Space Project with No.\,CMS-CSST-2021-A05. FAST is a Chinese national mega-science facility, operated by the National Astronomical Observatories of Chinese Academy of Sciences (NAOC). The Siena Galaxy Atlas was made possible by funding support from the U.S. Department of Energy, Office of Science, Office of High Energy Physics under Award Number DE-SC0020086 and from the National Science Foundation under grant AST-1616414. Funding for the Sloan Digital Sky Survey IV has been provided by the Alfred P. Sloan Foundation, the U.S. Department of Energy Office of Science, and the Participating Institutions. SDSS acknowledges support and resources from the Center for High-Performance Computing at the University of Utah. The SDSS web site is www.sdss4.org. SDSS is managed by the Astrophysical Research Consortium for the Participating Institutions of the SDSS Collaboration including the Brazilian Participation Group, the Carnegie Institution for Science, Carnegie Mellon University, Center for Astrophysics | Harvard \& Smithsonian (CfA), the Chilean Participation Group, the French Participation Group, Instituto de Astrofísica de Canarias, The Johns Hopkins University, Kavli Institute for the Physics and Mathematics of the Universe (IPMU) / University of Tokyo, the Korean Participation Group, Lawrence Berkeley National Laboratory, Leibniz Institut für Astrophysik Potsdam (AIP), Max-Planck-Institut für Astronomie (MPIA Heidelberg), Max-Planck-Institut für Astrophysik (MPA Garching), Max-Planck-Institut für Extraterrestrische Physik (MPE), National Astronomical Observatories of China, New Mexico State University, New York University, University of Notre Dame, Observatório Nacional / MCTI, The Ohio State University, Pennsylvania State University, Shanghai Astronomical Observatory, United Kingdom Participation Group, Universidad Nacional Autónoma de México, University of Arizona, University of Colorado Boulder, University of Oxford, University of Portsmouth, University of Utah, University of Virginia, University of Washington, University of Wisconsin, Vanderbilt University, and Yale University.

%%%%%%%%%%%%%%%%%%%%%%%%%%%%%%%%%%%%%%%%%%%%%%%%%%%%%%%
%%% Conflict of interest. ????????????
%%%%%%%%%%%%%%%%%%%%%%%%%%%%%%%%%%%%%%%%%%%%%%%%%%%%%%%
\InterestConflict{The authors declare that they have no conflict of interest.}

%%%%%%%%%%%%%%%%%%%%%%%%%%%%%%%%%%%%%%%%%%%%%%%%%%%%%%%
%%% Supplements. ????????, ????
%%%%%%%%%%%%%%%%%%%%%%%%%%%%%%%%%%%%%%%%%%%%%%%%%%%%%%%
%\Supplements{}

%%%%%%%%%%%%%%%%%%%%%%%%%%%%%%%%%%%%%%%%%%%%%%%%%%%%%%%
%%% Reference section. ?ο?????
%%% citation in the content using "some words~ \citep{1,2}".
%%% ~ is needed to make the reference number is on the same line with the word before it.
%%%%%%%%%%%%%%%%%%%%%%%%%%%%%%%%%%%%%%%%%%%%%%%%%%%%%%%
{\small \setlength{\baselineskip}{-1pt}
\bibliographystyle{raa}
\bibliography{references}
}

\end{multicols}

\end{document}